\documentclass[11pt]{article}

\usepackage[T1]{fontenc}

\usepackage{amsmath,amssymb}
\usepackage{amsthm}
\usepackage[active]{srcltx}
\usepackage{xcolor}

\newcommand{\R}{\mathbb{R}}
\newcommand{\C}{\mathbb{C}}

\newcommand{\LL}{{\cal L}}

\newcommand{\scal}[2]{\langle #1| #2\rangle}

\newcommand{\eps}{\epsilon}
\newcommand{\veps}{\varepsilon}

\newcommand{\ot}{\otimes}

\newcommand{\bld}[1]{\boldsymbol{#1}}

\newcommand{\we}{\wedge}
\newcommand{\un}[1]{\underline{#1}}
\newcommand{\bth}{\boldsymbol{\theta}}
\newcommand{\vth}{\vec{\theta}}
\newcommand{\lr}{\lrcorner}
\newcommand{\ubth}{\underline{\boldsymbol{\theta}}}
\newcommand{\bed}{\boldsymbol{d}}

\newcounter{mnotecount}[section]

\numberwithin{equation}{section}
\numberwithin{thr}{section}

\begin{document}

\title{ADM-like Hamiltonian formulation of gravity in the teleparallel geometry\footnote{This is an author-created version of a paper accepted for publication in {\em Gen. Rel. Grav.}}}
\author{ Andrzej Oko{\l}\'ow}
\date{September 19, 2013}

\maketitle
\begin{center}
{\it  Institute of Theoretical Physics, Warsaw University\\ ul. Ho\.{z}a 69, 00-681 Warsaw, Poland\smallskip\\
oko@fuw.edu.pl}
\end{center}
\medskip

\begin{abstract}
We present a new Hamiltonian formulation of the Teleparallel Equivalent of General Relativity (TEGR) meant to serve as the departure point for canonical quantization of the theory. TEGR is considered here as a theory of a cotetrad field on a spacetime. The Hamiltonian formulation is derived  by means of an ADM-like $3+1$ decomposition of the field and without any gauge fixing. A complete set of constraints on the phase space and their algebra are presented. The formulation is described in terms of differential forms.  
\end{abstract}

\section{Introduction}

Among current approaches \cite{app,carlip} to quantum gravity there is no one based on the Teleparallel Equivalent of General Relativity (TEGR) (see \cite{mal-rev} for the latest review on the theory). Therefore it is worth to check whether it is possible to quantize gravity in this formulation. Our project is to check whether it is possible to quantize TEGR {in a background independent (diffeomorphism invariant) manner} by means of the method of canonical quantization or, if necessary, a modification of this method. 

As the departure point for canonical quantization of TEGR we would like to use a canonical formulation of the theory satisfying the following conditions:
\begin{enumerate}
\item the formulation is derived without any gauge fixing;
\item the canonical variables are a cotetrad field restricted to a space-like slice of the spacetime and the momentum conjugate to it;
\item the complete set of constraints is known as well as its division into the first and second class constraints;
\item the formulation is of the ADM-type, i.e. the non-dynamical degrees of freedom of the configuration variables are parameterized by the lapse function and the shift vector field (in the formulation the two latter variables play a role of Lagrangian multipliers).
\end{enumerate}
\noindent Let us now justify these requirements. 

Condition 1 corresponds to our wish to construct a quantum model of TEGR possessing as many symmetries of the classical theory as possible {including (spatial) diffeomorphism invariance}. 

Regarding Condition 2 let us emphasize that from the point of view of canonical formalism TEGR is a constrained system \cite{bl,nester,wall-av,maluf-2,maluf-1,maluf}. Therefore when quantizing canonically the theory we have to choose one of the following two strategies: $(i)$ ``first quantize, then solve the constraints'' (this is the Dirac strategy) or $(ii)$ ``first solve the constraints, then quantize''. Since we are unable to solve the constraints classically we have to choose the first strategy, which means in particular that the first step of the quantization is a construction of kinematic quantum states (here the adjective ``kinematic'' emphasizes the fact that these quantum states correspond to all classical states in the phase space of the theory, that is, to states which satisfy and states which do not satisfy the constraints). On the other hand at the Lagrangian level TEGR can be described either $(i)$ as a theory of a cotetrad field and a Lorentz connection of zero curvature---see e.g. \cite{obh,mielke,kopcz,nest-kop} or $(ii)$ as a theory of a cotetrad field only---see e.g. \cite{waldyr,mal-rev}. Thus the construction of the kinematic quantum states can be based on canonical variables derived either $(i)$ from the cotetrad field and the Lorentz connection or $(ii)$ the cotetrad field only. However, by now there is no method of constructing quantum states for a theory of a connection with a {\em non-compact} structure group (see \cite{oko-H,oko-ncomp})---one consequence of this fact is that the kinematic Hilbert space of Loop Quantum Gravity (LQG) \cite{al-hoop} is based on the real Ashtekar-Barbero connection \cite{barb} (the structure group of it is $SU(2)$) instead of the complex Ashtekar-Sen connection \cite{a-var-1,a-var-2} (the structure group of it is $SL(2,\C)$). Thus we are left with the second possibility expressed as Condition 2.

Constraints on the phase space have to be incorporated in a way into the structure of a resulting quantum model. Therefore one should know a complete set of the constraints. Moreover, at the quantum level one usually treats first class constraints in a different way than second class ones. Therefore one should know which constraints are of the first class and which are of the second class. This justifies Condition 3.

Condition 4 was imposed because of our wish to {quantize TEGR in a diffeomorphism invariant manner and, in particular, to} apply some ideas developed in LQG which is {a diffeomorphism invariant model of quantum gravity} based on an ADM-like Hamiltonian formulation of General Relativity (GR) (see e.g. review papers \cite{rev,rev-1}). First of all, an ADM-like formulation of GR provides a vector constraint which generates on the phase space gauge transformations corresponding to spatial diffeomorphisms. In particular, this fact was used in LQG to ``solve'' the vector constraint by finding quantum states invariant with respect to an action of spatial diffeomorphisms (see e.g. \cite{cq-diff}). Moreover, in recent years there were constructed two very interesting quantum models of gravity coupled to a matter field: in \cite{kg-tt} gravity is coupled to a dust and in \cite{lewand} to a scalar field. These models combine the standard LQG methods with so called relational observables \cite{rov,ditt} and underlying canonical formulations of GR coupled with matter fields \cite{kuchar,kuchar-1} are of the ADM-type.

In this paper we present a Hamiltonian formulation of TEGR satisfying all Conditions. The formulation was derived from the following action of TEGR {\cite{kopcz,thir,nest-kop,wall-av,waldyr,wall-act}}:
\begin{equation}
S[\bth^A]=\int-\frac{1}{2}(\bed\bth^A\we\bth_B)\we\star(\bed\bth^B\we\bth_A)+\frac{1}{4}(\bed\bth^A\we\bth_A)\we\star(\bed\bth^B\we\bth_B).
\label{act}
\end{equation}
In this action $(\bth^A)$ $(A=0,1,2,3)$ is a cotetrad field on a four-dimensional manifold i.e. $(\bth^A)$ is a collections of differential one-forms which are linearly independent at every point of the manifold, $\bed$ is the exterior derivative of differential forms on the manifold and $\star$ is the Hodge operator defined by a Lorentzian metric given by the cotetrad $(\bth^A)$. To describe the resulting Hamiltonian formulation we used a special kind of canonical formalism {adapted to differential forms patterned on that described in \cite{ham-diff,wall-av,mielke}}. 

The Hamiltonian formulation obtained form \eqref{act} is well defined. In this paper we present a Hamiltonian, a complete set of constraints on the phase space and a constraint algebra. To the best of our knowledge this is the first Hamiltonian formulation of TEGR satisfying Conditions 1, 2 and 3 which was derived by means of an ADM-like $3+1$ decomposition of the cotetrad field. According to this formulation  TEGR is a constrained system with first class constraints only. Among gauge transformations generated on the phase space by the constraints one can identify action of spatial diffeomorphisms generated by a vector constraint and local Lorentz transformations defined by some other constraints---it is worth to note that the Lorentz transformations act on the canonical variables in a non-standard way.  

{Taking advantage of} these results we proceeded further with canonical quantization of TEGR and carried out the first step of the Dirac procedure, that is, we constructed a space of kinematic quantum states for TEGR. This construction will be published soon in a series of papers \cite{q-stat,q-suit,ham-nv,q-tegr} which currently are in preparation.

The paper is organized as follows: after preliminaries (Section 2) we present in Section 3 the Hamiltonian description of TEGR, that is, a description of the phase space, a Hamiltonian, a complete set of constraints on the phase space and a constraint algebra. Section 3 ends by a discussion {of the results and a comparison with earlier works \cite{wall-av} and \cite{maluf}}. Next, in Section 4 we carry out the Legendre transformation and derive the Hamiltonian and the constraints (a derivation of the constraint algebra will be presented in an accompanying paper \cite{oko-tegr-II}). Let us emphasize that the derivation of the Hamiltonian and the constraints is rather long and technically complicated. Therefore we reversed the usual order of the presentation: we placed the results and the discussion right after preliminaries for the sake of readers not interested in the derivation and placed the derivation in the last section of the paper (Section 4) which plays a role of a technical appendix.

\section{Preliminaries}

Let $\mathbb{M}$ be a four-dimensional oriented vector space equipped with a scalar product $\eta$ of signature $(-,+,+,+)$. We fix an orthonormal basis $(v_A)$ $(A=0,1,2,3)$ such that the components $(\eta_{AB})$ of $\eta$ given by the basis form a matrix ${\rm diag}(-1,1,1,1)$. The matrix $(\eta_{AB})$ and its inverse $(\eta^{AB})$ will be used to, respectively, lower and raise capital Latin letter indeces.     

Let $\cal M$ be a four-dimensional oriented manifold. We assume that there exists a smooth  map $\bth:T{\cal M}\to \mathbb{M}$ such that for every $y\in{\cal M}$ the restriction of $\bth$ to the tangent space $T_y{\cal M}$ is a {\em linear isomorphism} between the tangent space and $\mathbb{M}$ which preserves the orientations. The map $\bth$ can be expressed by means of the orthogonal basis $(v_A)$ as
\[
\bth=\bth^A\ot v_A,
\]           
where $(\bth^A)$ are one-forms on $\cal M$. It is clear that the one-forms $(\bth^A)$ form a {\em coframe} or a {\em cotetrad field} on the manifold. 

The map $\bth$ can be used to pull back the scalar product $\eta$ on $\mathbb{M}$ to the manifold $\cal M$ turning thereby the manifold into a {\em spacetime}. We will denote the resulting Lorentzian metric by $g$,
\begin{equation}
g:=\eta_{AB}\bth^A\ot\bth^B.
\label{g}
\end{equation}
The metric $g$ defines a volume form $\bld{\eps}$ on $\cal M$ and a Hodge dual operator $\star$ mapping differential $k$-forms to $(4-k)$-forms on the manifold $(k=0,1,2,3,4)$. 

\subsection{TEGR}

In this paper we will treat TEGR as a theory of cotetrad fields on $\cal M$ which means that the configuration space of the theory will be a set of all the maps $\bth$ which satisfy the assumptions listed above. We choose the action \eqref{act} as one describing the dynamics of TEGR (for different but equivalent actions see e.g. \cite{obh,mielke}). Let us emphasize that the Hodge operator $\star$ appearing in \eqref{act} is given by the metric \eqref{g} and therefore it is a function of $(\bth^A)$. 

The passage from the action \eqref{act} to a Hamiltonian formulation requires as its first step a $3+1$ decomposition of: the manifold  $\cal M$, differential forms on it and a cotetrad $(\bth^A)$.

\subsection{$3+1$ decomposition of $\cal M$ \label{3+1-M}}

To carry out a $3+1$ decomposition of the action \eqref{act} we have to impose some additional assumptions on the manifold $\cal M$ and the map $\bth$. We require that 
\begin{enumerate}
\item ${\cal M}=\R\times\Sigma$, where $\Sigma$ is a three-dimensional manifold.
\item the map $\bth$ is such that for every $t\in\R$ the submanifold $\Sigma_t:=\{t\}\times\Sigma\subset{\cal M}$ is spatial with respect to $g$. 
\end{enumerate}

Assumption 1 allows us to introduce a family of curves in $\cal M$ parameterized by points of $\Sigma$---given $x\in\Sigma$ we define
\[
\R\ni t\mapsto (t,x)\in\R\times\Sigma={\cal M}.
\]
These curves generate a global vector field on $\cal M$ which will be denoted by $\partial_t$. 

Moreover, due to Assumption 1 there exists a function on $\cal M$ which maps a point $y$ to a number $\tau$  such that $y\in\Sigma_\tau$. Let us denote the function by $t$. Consider now a local coordinate frame $(x^i)$, $(i=1,2,3)$, on $\Sigma$. This coordinate frame together with the function $t$ define a local coordinate frame $(x^0\equiv t,x^i)\equiv(x^\mu)$ on $\cal M$.  Throughout the paper we will restrict ourselves to coordinate frames $(x^\mu)$ on $\cal M$ of this sort assuming additionally that each frame we are going to use is {\em compatible} with the orientation of the manifold. 

Note that the class of coordinate frames just introduced induces an orientation of $\Sigma$ which since now will be treated as an {\em oriented} manifold. 

Let us emphasize that in this paper the spacetime indeces will be denoted by lower case Greek letters and will range from $0$ to $3$ and the spatial indeces will be denoted by lower case Latin letters and will range from $1$ to $3$.    

A set of all cotetrad fields $(\bth^A)$ compatible with the orientation of $\cal M$ and satisfying Assumption 2 will be called {\em restricted configuration space} {and denoted by $\bld{\Theta}$}.

In order to not be troubled by boundary terms in the Hamiltonian formulation we assume that
\begin{enumerate}
\setcounter{enumi}{2}
\item $\Sigma$ is a compact manifold without boundary. 
\end{enumerate}

\subsection{Decomposition of differential forms \label{dec-forms}}

Denote by $\bed$ the exterior derivative of forms on $\cal M$ and by $d$ the exterior derivative of forms on $\Sigma$. A $k$-form $\alpha$ on $\cal M$ can be decomposed with respect to the decomposition ${\cal M}=\R\times\Sigma$ as follows {\cite{ham-diff,wall-av,mielke}} 
\[
\alpha={}^\perp\alpha+\un{\alpha},
\]    
where
\[
{}^\perp\alpha:=\bed t\we \partial_t\lr\alpha 
\]
is its ``time-like'' part and
\[
\un{\alpha}:=\partial_t\lr(\bed t\we\alpha)
\]
its ``space-like'' part. It is convenient to denote
\[
\alpha_\perp\equiv\partial_t\lr\alpha.
\]
Then 
\[
{}^\perp\alpha=\bed t\we \alpha_\perp.
\]

$\un{\alpha}$ is a $k$-form on $\cal M$ which naturally defines a family $\{\un{\alpha}_t\}_{t\in\R}$ of $k$-forms on $\Sigma$: if $\varphi_t:\Sigma\mapsto\Sigma_t\subset{\cal M}$ denotes the natural embedding then
\[
\un{\alpha}_t:=\varphi^*_t\un{\alpha}.
\]  
Moreover, it is possible to restore the original form $\un{\alpha}$ from the family $\{\un{\alpha}_t\}_{t\in\R}$: given the latter one we define
\begin{align*}
&\partial_t\lr\un{\alpha}:=0, &&\un{\alpha}(\vec{X}_1,\ldots,\vec{X}_k):=\un{\alpha}_t(\vec{X}_1,\ldots,\vec{X}_k) 
\end{align*}
where all vector fields $(\vec{X}_1,\ldots,\vec{X}_k)$ are tangent to the submanifold $\Sigma_t$. Therefore in the sequel we will not distinguish between $\un{\alpha}$ and the forms {$\{\un{\alpha}_t\}_{t\in\R}$}. 

There is however a subtlety concerning Lie derivatives of forms $\un{\alpha}$ and $\un{\alpha}_t$. Let $\vec{X}$ be a vector field on $\cal M$ tangent to the foliation $\{\Sigma_{t}\}_{t\in\R}$. Denote by $\LL_{\vec{X}}$ the Lie derivative on $\cal M$ with respect to $\vec{X}$  and by $\LL^t_{\vec{X}}$  the Lie derivative on $\Sigma$ with respect to {$\varphi^{-1}_{t*}\vec{X}$}. Then in general $\LL_{\vec{X}}\un{\alpha}$ cannot be identified with the family $\{\LL^t_{\vec{X}}\un{\alpha}_t\}_{t\in\R}$. Indeed, if $\alpha$ is for example a one-form on $\cal M$ then    
\[
\LL_{\vec{X}}\un{\alpha}=(X^\mu\partial_\mu\un{\alpha}_\nu+\un{\alpha}_\mu\partial_\nu X^\mu)\bed x^\mu=\un{\alpha}_i\partial_0 X^i \bed t+(X^i\partial_i\un{\alpha}_j+\un{\alpha}_i\partial_j X^i)\bed x^j.
\]
and only the last term in this equation can be identified with the family $\{\LL^t_{\vec{X}}\un{\alpha}_t\}_{t\in\R}$. However, in the sequel we will never encounter Lie derivatives $\LL_{\vec{X}}\un{\alpha}$ as defined above but we will do encounter derivatives of forms on $\Sigma$ with respect to a vector field on the manifold. Since we would like our notation to be as simple as possible since now we will use the symbol $\LL_{\vec{X}}\un{\alpha}$ to denote the derivative $\LL^t_{\vec{X}}\un{\alpha}_t$.

Similarly, $\alpha_\perp$ is a form on $\cal M$, but it can be treated as a {one parameter family $\{\alpha_{\perp t}\}_{t\in\R}$ of forms  on $\Sigma$ defined by pull-back with respect the natural embeddings of $\Sigma$ into $\cal M$}. {Consequently, the $k$-form $\alpha$ on $\cal M$ can be identified with a family $\{\alpha_{\perp t},\un{\alpha}_t\}_{t\in\R}$ of, respectively, $(k-1)$-forms and $k$-forms on $\Sigma$. It is easy to see that this $3+1$ decomposition of forms is equivalent to the standard decomposition carried out with respect to a coordinate frame $(t,x^i)$ adapted to the decomposition ${\cal M}=\R\times \Sigma$.}    

Basic properties of the maps $\alpha\mapsto{}^\perp\alpha$, $\alpha\mapsto\alpha_\perp$ and $\alpha\mapsto\un{\alpha}$ read {\cite{ham-diff,wall-av,mielke}}:
\begin{equation}
\begin{aligned}
&{}^\perp({}^\perp\alpha)={}^\perp\alpha, & &\un{({}^\perp\alpha)}={}^\perp(\un{\alpha})=0,  & &\un{(\un{\alpha})}=\un{\alpha},\\
& \un{(\alpha\we\beta)}=\un{\alpha}\we\un{\beta}, & & {}^\perp(\alpha\we\beta)=({}^\perp\alpha)\we\un{\beta}+\un{\alpha}\we({}^\perp\beta), & & \un{\alpha_\perp}=\alpha_\perp,\\
& \partial_t\lr\un{\alpha}=0, & & (\alpha\we\beta)_\perp=\alpha_\perp\we\un{\beta}+(-1)^k\un{\alpha}\we\beta_\perp, & & \un{(\bed \alpha)}=d\un{\alpha}, \\
& (\bed\alpha)_\perp=\LL_{\partial_t}\un{\alpha}-d\alpha_\perp, & & \bed \alpha=\bed t\we\LL_{\partial_t}\un{\alpha}-\bed t\we d\alpha_\perp+d\un{\alpha},    & & 
\end{aligned}
\label{perp-un}  
\end{equation}
In these formulae $\alpha$ is a $k$-form and $\LL_{\partial_t}$ denotes the Lie derivative with respect to the vector field $\partial_t$. {Let us note that there is a slight difference between the formulae above and their counterparts in \cite{ham-diff,wall-av}: here we use the exterior derivative $d$ acting on differential forms on $\Sigma$ while in these papers the corresponding derivative $\un{d}$ acts on forms defined on $\cal M$.}

\subsection{Decomposition of the cotetrad}

Since each $\bth^A$ is a one-form it decomposes as
\begin{equation}
\bth^A=\bth^A_\perp \bed t + \ubth^A.
\label{corep-dec}
\end{equation}
It turns out that $\bth^A_\perp$ is a function of $\ubth^A$ and some additional parameters \cite{nester,os}:
\begin{align}
&\bth^A_\perp=N\xi^A+\vec{N}\lr\un{\bth}^A, 
\label{theta-0-xi-a}\\
&\xi^A:=-\frac{1}{3!}\veps^A{}_{BCD}*(\un{\bth}^B\we\un{\bth}^C\we\un{\bth}^D)
\label{xi-a},\\
&N>0, \label{N>0}
\end{align} 
where
\begin{enumerate}
\item $N$ is a function on $\cal M$ called {\em lapse};  
\item $\vec{N}$ is a vector field on $\cal M$ called {\em shift}. It is tangent at each point to a submanifold $\Sigma_{t}$ passing through the point---in an admissible coordinate system $(t,x^i)$ 
\[
\vec{N}=N^i\partial_i;
\]   
\item $\veps_{ABCD}$ is a volume form on $\mathbb{M}$ given by the scalar product $\eta$;
\item $*$ is the Hodge operator on $\Sigma_{t}$ given by a Riemannian metric $q$ {induced on $\Sigma_t$ by $g$}:     
\begin{equation}
\begin{gathered}
q=q_{ij}dx^i\ot dx^j:=g_{ij}dx^i\ot dx^j=\eta_{AB}\ubth^A\ot\ubth^B,\\
q_{ij}=\eta_{AB}\bth^A_i\bth^B_j=\eta_{AB}\ubth^A_i\ubth^B_j.
\end{gathered}
\label{q}
\end{equation}      
\end{enumerate}
The functions $\xi^A$ satisfy the following important conditions \cite{nester}: 
\begin{align}
\xi^A\xi_A&=-1, &\xi^A\un{\bth}_A&=0.
\label{xixi}
\end{align}
These two equations imply
\begin{align}
\xi^Ad\xi_A&=0, &d\xi^A\we\un{\bth}_A+\xi^Ad\un{\bth}_A&=0.
\label{dxixi}
\end{align}

Fixing the value of the index $\mu$ we can treat the four components $\bth^A_\mu$ as a function on $\cal M$ valued in $\mathbb{M}$. The conditions \eqref{xixi} mean that for every $y\in{\cal M}$ the vectors $(\xi^A(y),\bth^A_i(y))$ form a basis of $\mathbb{M}$.

The decomposition of the cotetrad allows us to change the way we parameterize the restricted configuration space---instead of $(\bth^A_\perp,\un{\bth}^A)$ we will use $(N,\vec{N},\un{\bth}^A)$ as parameters on this space. This change is obviously motivated by our wish to obtain an ADM-like Hamiltonian formulation of TEGR and can be seen as a source of difference between this approach and that of \cite{maluf-1,maluf}---see Section \ref{sec-comp} for a comparison between these two approaches. 

\subsection{Decomposition of the spacetime metric}

Setting to \eqref{g} the cotetrad $(\bth^A)$ decomposed according to \eqref{corep-dec} and \eqref{theta-0-xi-a} we obtain the standard $3+1$ decomposition of the spacetime metric $g$ \cite{adm}:  
\begin{equation}
g=(-N^2+N^iN^jq_{ij})\,\bed t^2+2N^iq_{ij}\,\bed t\bed x^j+q,
\label{g-dec}
\end{equation}
where $q$ given by \eqref{q} is the Riemannian metric induced on $\Sigma_t$. This decomposition justifies calling the function $N$ the lapse and the vector field $\vec{N}$ the shift (for a more precise justification see \cite{os}). 

 The metric $q$ and its inverse $q^{-1}$,
\begin{equation}
q^{-1}:=q^{ij}\partial_i\ot\partial_j, \ \ \ q^{ij}q_{jk}=\delta^i{}_k,
\label{q-1}
\end{equation}
will be used to, respectively, lower and raise, indeces (here: lower case Latin letters) of components of tensor fields defined on $\Sigma$. In particular we will often map one-forms to vector fields on $\Sigma$---a vector field corresponding to a one form $\alpha$ will be denoted by $\vec{\alpha}$ i.e. if (locally) $\alpha=\alpha_i dx^i$ then
\[
\vec{\alpha}:=q^{ij}\alpha_i\partial_j.
\]    

The metric $q$ defines a volume form $\eps$ on $\Sigma$ and the Hodge operator $*$ acting on differential forms on the manifold.  

Let us emphasize finally that (as it follows from \eqref{q}) the metric $q$ can be defined explicitely in terms of the restricted forms $(\un{\bth}^A)$. Therefore all object defined by $q$ (as $q^{-1}$, $\eps$ and $*$) are in fact functions of $(\un{\bth}^A)$.      

\section{Hamiltonian description of TEGR}

In this section we are going to present the canonical framework of TEGR derived from the action \eqref{act}. Let us emphasize that to describe the framework we will use the Hamiltonian formalism adapted to differential forms \cite{ham-diff,mielke} (see also \cite{os}). 

Before we will show the results let us simplify the notation---since now we will denote the ``space-like'' part of the one-form $\bth^A$ by $\theta^A$, i.e.
\begin{equation}
\un{\bth}^A\equiv\theta^A
\label{bt-t}
\end{equation}
and its Lie derivative with respect to $\partial_t$ by $\dot{\theta}^A$ i.e.
\begin{equation}
\LL_{\partial_t}\un{\bth}^A \equiv \dot{\theta}^A.
\label{Ltbt}
\end{equation}

\subsection{Hamiltonian and constraints \label{H-c}}

In the action \eqref{act} there is no Lie derivative with respect to $\partial_t$ of the lapse $N$ and the shift $\vec{N}$ but there is one of $\theta^A$. Therefore the two former variables are treated as Lagrange multipliers, while the latter one as one of the canonical variables. A point in the phase space of the theory consists of
\begin{enumerate}
\item a quadruplet of one-forms $(\theta^{A})$ on $\Sigma$ such that a metric
\begin{equation}
q=\eta_{AB}\theta^A\ot\theta^B
\label{q-tt}
\end{equation}
on $\Sigma$ is {\em Riemannian} (i.e. {\em positive definite}); 
\item a quadruplet of two-forms $(p_A)$ on the manifold---$p_A$ is the momentum conjugate to $\theta^A$.
\end{enumerate}         
Equivalently, a point in the phase space of the theory consists of
\begin{enumerate}
\item a map $\theta:T\Sigma\to \mathbb{M}$ such that   for every $x\in\Sigma$ the restriction of $\theta$ to $T_x\Sigma$ is a linear map and the pull-back $\theta^*\eta$ of the scalar product $\eta$ is a Riemannian metric on $\Sigma$; 
\item the conjugate momentum $p$ as a two-form on $\Sigma$ valued in $\mathbb{M}^*$ being the dual space to $\mathbb{M}$.
\end{enumerate}      

The Legendre transformation is given by\footnote{For a definition of the partial derivative ${\partial L_\perp}/{\partial {\dot{\theta}}^A}$ see \cite{os}.}    
\begin{equation}
p_A=\frac{\partial L_\perp}{\partial {\dot{\theta}}^A},
\label{leg-tr-0}
\end{equation}
where $L$ is the integrand in \eqref{act}. The momentum turns out to be quite complicated function of the variables $N,\vec{N},\theta^A$ and $\dot{\theta}^A$:
\begin{multline}
p_A=N^{-1}\Big({\theta}_B\we{*}[{\dot{\theta}}^B\we{\theta}_A-N({d}\xi^B\we{\theta}_A-{d}{\theta}^B\we\xi_A)-{\cal L}_{\vec{N}}{\theta}^B\we{\theta}_A ]-\\-\frac{1}{2}{\theta}_A\we{*}[{\dot{\theta}}^B\we{\theta}_B-N({d}\xi^B\we{\theta}_B-{d}{\theta}^B\we\xi_B)-{\cal L}_{\vec{N}}{\theta}^B\we{\theta}_B]\Big),
\label{p_A-exp}
\end{multline}
where $\LL_{\vec{N}}$ denotes the Lie derivative on $\Sigma_t$ with respect to $\vec{N}$.

The Legendre transformation is not invertible and one encounters the following {\em primary constraints}
\begin{align}
&\theta^A\we*d\theta_A+\xi^Ap_A=0,\label{B-constr}\\
&\theta^A\we*p_A-\xi^Ad\theta_A=0 \label{R-constr}
\end{align}
called here {\em boost} and {\em rotation} constraints respectively (for a justification of the names see Section \ref{gauge}). Their smeared versions read
\begin{align}
B(a):=&\int_\Sigma a\we(\theta^A\we*d\theta_A+\xi^Ap_A),\label{B-sm}\\
R(b):=&\int_\Sigma b\we(\theta^A\we*p_A-\xi^Ad\theta_A),\label{R-sm}
\end{align}
where $a$ and $b$ are one-forms on $\Sigma$.
   
The Hamiltonian 
\[
H_0:=\int_\Sigma \dot{\theta}^A\we p_A-L_\perp
\]
is unambiguously defined on the image of the Lagrange transformation (that is, on a subset of the phase space distinguished by vanishing of the primary constraints) and is of the following form
\begin{multline}
H_0[{\theta}^A,p_B,N,\vec{N}]=\int_\Sigma N\Big(\frac{1}{2}(p_A\we\theta^B)\we*(p_B\we\theta^A)-\frac{1}{4}(p_A\we\theta^A)\we*(p_B\we\theta^B)-\\-\xi^A\we{d}p_A+\frac{1}{2}(d\theta_A\we\theta^B)\we{*}(d\theta_B\we\theta^A)-\frac{1}{4}(d\theta_A\we\theta^A)\we{*}(d\theta_B\we\theta^B)\Big)-\\-{d}{\theta}^A\we(\vec{N}\lrcorner p_A)-(\vec{N}\lrcorner{\theta}^A)\we {d}p_A.
\label{H_0}
\end{multline}
It can be extended to the whole phase space by adding the primary constraints:
\begin{equation}
H[{\theta}^A,p_B,N,\vec{N},a,b]=H_0[\theta^A,p_B,N,\vec{N}]+B(a)+R(b),
\label{full-ham}
\end{equation}
where the one-forms $a$ and $b$ play the role of Lagrange multipliers.
 
The Lagrange multipliers $N$ and $\vec{N}$ appearing in the Hamiltonian \eqref{full-ham} generate the following {\em secondary constraints} 
\begin{equation}
\begin{aligned}
&\frac{1}{2}(p_A\we\theta^B)\we*(p_B\we\theta^A)-\frac{1}{4}(p_A\we\theta^A)\we*(p_B\we\theta^B)-\xi^A\we{d}p_A+\\+&\frac{1}{2}(d\theta_A\we\theta^B)\we{*}(d\theta_B\we\theta^A)-\frac{1}{4}(d\theta_A\we\theta^A)\we{*}(d\theta_B\we\theta^B)=0,\\
&-{d}{\theta}^A\we(\partial_i\lrcorner p_A)-(\partial_i\lrcorner{\theta}^A)\we {d}p_A=0.
\end{aligned}  
\label{sec-constr}
\end{equation}           
called {\em scalar} and {\em vector} constraints respectively. Smeared versions of the constrains read
\begin{align}
S(M):=&\int_\Sigma M\Big(\frac{1}{2}(p_A\we\theta^B)\we*(p_B\we\theta^A)-\frac{1}{4}(p_A\we\theta^A)\we*(p_B\we\theta^B)-\xi^A\we{d}p_A+\nonumber\\+&\frac{1}{2}(d\theta_A\we\theta^B)\we{*}(d\theta_B\we\theta^A)-\frac{1}{4}(d\theta_A\we\theta^A)\we{*}(d\theta_B\we\theta^B)\Big),\label{S-sm}\\
V(\vec{M}):=&\int_\Sigma -{d}{\theta}^A\we(\vec{M}\lrcorner p_A)-(\vec{M}\lrcorner{\theta}^A)\we {d}p_A,\label{V-sm}
\end{align}     
where $M$ is a function on $\Sigma$ and $\vec{M}$ a vector field on the manifold.     

The Hamiltonian $H_0$ is a sum of the smeared scalar and vector constraints,
\begin{equation}
H_0[{\theta}^A,p_B,N,\vec{N}]=S(N)+V(\vec{N}),
\label{H0-SV}
\end{equation}
and the extended Hamiltonian is a sum of all the constraints:
\begin{equation}
H[{\theta}^A,p_B,N,\vec{N},a,b]=S(N)+V(\vec{N})+B(a)+R(b).
\label{full-ham-c}
\end{equation}

\subsection{Constraint algebra \label{c-alg}}

In this section we present the algebra of constraints derived in the accompanying paper \cite{oko-tegr-II}. The Poisson brackets of the smeared boosts and rotation constrains read:
\begin{equation}
\begin{aligned}
&\{B(a),B(a')\}=-R(*(a\we a')),\\
&\{R(b),R(b')\}=R(*(b\we b')),\\
&\{B(a),R(b)\}=B(*(a\we b)).
\end{aligned}
\label{BRBR}
\end{equation}
The bracket of the scalar constraints is most complex:
\begin{multline*}
\{S(M),S(M')\}=V(\vec{m})+B\Big(\theta^B*(m\we p_B)-\frac{1}{2}*(m\we\xi^B*d\theta_B)-\\-*[m\we*(\theta^B\we*p_B)]-\frac{1}{2}*(*m\we\theta^B)*p_B+\frac{1}{2}*[*(m\we\theta^B)\we*p_B]\Big)+\\+R\Big(-\theta^B*(m\we d\theta_B)-\frac{1}{2}*(m\we\xi^B*p_B)+\\+*[m\we*(\theta^B\we*d\theta_B)]+\frac{1}{2}*(*m\we\theta^B)*d\theta_B-\frac{1}{2}*[*(m\we\theta^B)\we*d\theta_B]\Big)
\end{multline*}
where 
\begin{equation}
m:=MdM'-M'dM.
\label{m}
\end{equation}
The brackets of  the boost and rotation constraints and the scalar one: 
\begin{align}
\{B(a),S(M)\}=&-B\Big(M[\theta^B*(p_B\we a)-\frac{1}{2}a*(p_B\we\theta^B)+d\xi_B*(a\we*\theta^B)]\Big)+\nonumber\\&+R\Big(*(dM\we a)\Big),\label{BS-res}\\
\{R(b),S(M)\}=&-R\Big(M[\theta^B*(p_B\we b)-\frac{1}{2}b*(p_B\we\theta^B)+d\xi_A*(b\we*\theta^A)]\Big)-\nonumber\\&-B\Big(*(dM\we b)\Big).\label{RS-res}
\end{align}
The brackets of the vector constrains:
\begin{equation}
\begin{aligned}
\{V(\vec{M}),V(\vec{M}')\}=&V(\LL_{\vec{M}}\vec{M}')\equiv V([\vec{M},\vec{M}']),\\
\{V(\vec{M}),S(M)\}=&S(\LL_{\vec{M}}M),\\
\{V(\vec{M}),B(a)\}=&B(\LL_{\vec{M}}a),\\
\{V(\vec{M}),R(b)\}=&R(\LL_{\vec{M}}b),
\end{aligned} 
\label{VV}
\end{equation} 
where $\LL_{\vec{M}}$ denotes the Lie derivative on $\Sigma$ with respect to the vector field $\vec{M}$.  

Thus the Poisson bracket of any pair of the constraints $S(M)$, $V(\vec{M})$, $B(a)$ and $R(b)$ is a combination of the constraints. Since the Hamiltonian \eqref{full-ham-c} is a sum of the constraints each of the constraints listed above is preserved by the time evolution hence the list of the constraints is {\em complete}. All these mean that the constraints are of the {\em first class}. Note, however, that the constraint algebra is not a Lie algebra---most of the Poisson brackets are combinations of the constraints smeared with fields being functions of the canonical variables. 

\subsection{Discussion}

The main conclusion is that the Legendre transformation applied to the action \eqref{act} as a functional of cotetrad fields leads to a well defined ADM-like Hamiltonian formulation of TEGR. It is a constrained Hamiltonian system with first class constraints only. As a consequence of parameterizing the ``time-like'' part $\bth^A_\perp$ of the cotetrad by means of the lapse $N$ and the shift $\vec{N}$ (see Equation \eqref{theta-0-xi-a}) there appear in this formulation the scalar and the vector constraints.  

{Regarding the action \eqref{act}, its integrand $L$ differs from the integrand  $L_{HE}$ of the Hilbert-Einstein action for GR---note that $L$ contains only first derivatives of $(\bth^A)$, while $L_{HE}$ is known to contain second derivatives of a metric. Since the metric is an algebraic function of $(\bth^A)$ both integrands have to differ by an exact four-form containing second derivatives of the cotetrad field \cite{thir}:
\[
L_{EH}=L+\bed(\bth_A\we\star \bed\bth^A).
\]    
Of course, a derivation of a Hamiltonian formulation of TEGR from the r.h.s. of this equation would be more complicated since then we would have to deal with second derivatives of the cotetrad field. It is too difficult to predict how the Hamiltonian and the constraints would change if we kept the exact form, perhaps then a quite simple relation between the action \eqref{act} and the scalar constraint \eqref{S-sm} described in Section \ref{str-sc} would be lost.} 

{Let us also comment on Assumption 3 of Section \ref{3+1-M} which states that $\Sigma$ is a compact manifold without boundary. Such an assumption is often encountered in works concerning canonical quantization (see e.g. \cite{rev,rev-1}) but for other purposes is too restrictive. A comprehensive analysis  of boundary terms including non-Dirichlet boundary conditions (see \cite{kpt-0,kpt}) in the case of $\Sigma$ with boundary $\partial\Sigma $ would exceed the scope of this paper. Let us only remark that omitting Assumption 3 and imposing the Dirichlet boundary conditions (which usually is done tacitly) one obtains a boundary term in the Hamiltonian \eqref{full-ham-c} which originates from exact three-forms  on $\Sigma$ neglected in the derivation of the Hamiltonian (see a paragraph just above Equation \eqref{ham-expl}). The boundary term reads 
\[
\int_{\partial \Sigma} (N\xi^A+\vec{N}\lr\theta^A)p_A.
\]   
} 

\subsubsection{Gauge transformations \label{gauge}}

Since the action \eqref{act} is invariant with respect to (orientation preserving) diffeomorphisms of $\cal M$ one can expect that there exist gauge transformations on the phase space of the Hamiltonian formulation generated by (orientation preserving) diffeomorphisms of the slice $\Sigma$. Moreover, as stated in \cite{waldyr}, the action is invariant with respect to local Lorentz transformations therefore there should exist corresponding gauge transformations on the phase space.       

As shown in \cite{os} the vector constraint \eqref{V-sm} can be alternatively expressed as 
\[
V(\vec{M})=\int_\Sigma p_A\we({\cal L}_{\vec{M}}{\theta}^A)=-\int_\Sigma{\theta}^A\we{\cal L}_{\vec{M}}p_A,
\]
hence we have
\begin{align*}
\{\theta^A,V(\vec{M})\}&={\cal L}_{\vec{M}}{\theta}^A, & \{p_A,V(\vec{M})\}&={\cal L}_{\vec{M}}p_A.
\end{align*}
This means that gauge transformations given by the vector constraint $V(\vec{M})$ coincide with pull-backs of the canonical variables generated by diffeomorphisms moving points along integral curves of the vector field $\vec{M}$.

Now let us show that local Lorentz transformations on the phase space are generated by the constraints $B(a)$ and $R(b)$. Note first that the Poisson brackets \eqref{BRBR} are related closely to the Lie brackets of the Lie algebra of the Lorentz group. Indeed, there exists a basis $(\beta^i,\rho^j)$ $(i,j=1,2,3)$ of the Lie algebra consisting of generators of boosts $(\beta^i)$ and of generators of rotations $(\rho^j)$ such that
\begin{align*}
[\beta^i,\beta^j]&=-\check{\eps}^{ijk}\rho^l\delta_{kl},& \ \ [\rho^i,\rho^j]&=\check{\eps}^{ijk}\rho^l\delta_{kl},& \ \ [\beta^i,\rho^j]&=\check{\eps}^{ijk}\beta^l\delta_{kl}, 
\end{align*}
where $\check{\eps}^{ijk}$ is an antisymmetric symbol such that $\check{\eps}^{123}=1$. Defining
\begin{align*}
&\beta(A):=\beta^iA_i, &&\rho(B):=\rho^iB_i, 
\end{align*}
we can rewrite the Lie brackets above in the following form
\begin{equation}
\begin{aligned}
{}[\beta(A),\beta(A')]&=-\rho(\check{*}(A\we A')),\\ 
[\rho(B),\rho(B')]&=\rho(\check{*}(B\we B')), \\ 
[\beta(A),\rho(B)]&=\beta(\check{*}(A\we B))
\end{aligned}
\label{brbr}
\end{equation}
---here we regard the numbers $(A_i)$ and $(B_i)$ as components of one-forms $A$ and $B$, respectively, on the vector space $\R^3$ equipped with the standard scalar product $\delta_{ij}$ and the Hodge operator $\check{*}$ defined by the product.    

Taking into account that the metric $q$ defining the Hodge operator in \eqref{BRBR} is Riemannian the close relations between \eqref{BRBR} and \eqref{brbr} becomes evident and we are allowed to conclude that the constraints $B(a)$ and $R(b)$ generate local Lorentz transformations of the canonical variables---$B(a)$ generates local boosts and $R(b)$ local rotations.      

This conclusion can be strengthen by showing explicitely that at each point $x$ of $\Sigma$ the primary constraints $B(a)$ and $R(b)$ define an action of the Lorentz group on a space of quadruplets $(\theta^A(x))$, where $(\theta^A)$ runs over all fields allowed by the description of the phase space placed at the beginning of Section \ref{H-c}. We thus fix $x\in \Sigma$ and till Equation \eqref{nst-L-tr} we will consider values of fields only at this $x$, however, in order to keep the notation as simple as possible we will not use any special symbols to distinguish between fields and their values at $x$ i.e. the value $\theta^A(x)$ will be denoted by $\theta^A$ etc. 

Consider then the following system of differential equations imposed on components $\theta^A_i$ given by a fixed basis $(dx^i)$ of $T^*_x\Sigma$:  
\begin{multline}
\frac{d\theta^A_i}{d\lambda}(\lambda)=\{\theta^A_i(\lambda),B(a(\lambda))+R(b(\lambda))\}=\Big(\frac{\delta B(a(\lambda))}{\delta p_A}\Big)_i+\Big(\frac{\delta R(b(\lambda))}{\delta p_A}\Big)_i=\\=a_i(\lambda)\xi^A(\lambda)+\eps^{jk}{}_i(\lambda)b_j(\lambda)\theta^A_k(\lambda).
\label{t-trans}
\end{multline}
The components $a_i(\lambda)$ and $b_i(\lambda)$ depend on $\lambda$ in an arbitrary way. On the other hand, $\theta^A_i(\lambda)$ defines via \eqref{q} a scalar product 
\begin{equation}
q_{ij}(\lambda)=\eta_{AB}\theta^A_i(\lambda)\theta^B_j(\lambda)
\label{q-l}
\end{equation}
on $T_x\Sigma$, which in turn defines a volume form $\eps_{ijk}(\lambda)$ on $T_x\Sigma$. By rising the first two indeces of the volume form by the inverse $q^{ij}(\lambda)$ we obtain the tensor $\eps^{jk}{}_i(\lambda)$ appearing at the r.h.s. of \eqref{t-trans}. The scalar product \eqref{q-l} defines also a Hodge operator $*_\lambda$ acting on forms on $T_x\Sigma$ which can be used to express explicitely the function 
\begin{equation}
\xi^A(\lambda):=-\frac{1}{3!}\veps^A{}_{BCD}*_\lambda({\theta}^B(\lambda)\we{\theta}^C(\lambda)\we \theta^D(\lambda))=-\frac{1}{3!}\veps^A{}_{BCD}\eps^{ijk}(\lambda){\theta}^B_i(\lambda){\theta}^C_j(\lambda)\theta^D_k(\lambda)
\label{xi-l}
\end{equation}
corresponding to $\theta^A_i(\lambda)$ (see \eqref{xi-a}).
 
The gauge transformations of $\theta^A$ defined by the constraints $B(a)$ and $R(b)$ are given by Equations \eqref{t-trans}. More precisely, if $\lambda\mapsto\theta^A_i(\lambda)$ is a solution of the equations with the initial condition 
\begin{equation}
\theta^A_i(0)=\bar{\theta}^A_i
\label{ini}
\end{equation}
then any value $\theta^A_i(\lambda)$ is a result of the transformations acting on $\bar{\theta}^A_i$. 

Now we fix the initial values \eqref{ini} and will consider only the corresponding solution of \eqref{t-trans}. Although Equations \eqref{t-trans} appear to be highly nonlinear the solution of the equations can be found by solving a system of linear differential equations. To show this we note first that the scalar product \eqref{q-l} does not depend on $\lambda$: 
\begin{multline*}
\frac{dq_{ij}}{d\lambda}=\eta_{AB}\frac{d\theta^A_i}{d\lambda}\theta^{b}_j+\eta_{AB}\theta^{A}_i\frac{d\theta^B_j}{d\lambda}=\\=\eta_{AB}(a_i\xi^A+\eps^{kl}{}_ib_k\theta^A_l)\theta^{B}_j+\eta_{AB}\theta^{A}_i(a_j\xi^B+\eps^{kl}{}_jb_k\theta^B_l)=\eps^{k}{}_{ji}b_k+\eps^{k}{}_{ij}b_k=0.
\end{multline*}
In other words, the scalar product is constant along the solution and is a function of the initial values $\bar{\theta}^A_i$:
\[
q_{ij}(\lambda)=\eta_{AB}\bar{\theta}^A_i\bar{\theta}^B_j\equiv \bar{q}_{ij}.
\]
Consequently, analogous statements are true for all objects constructed from the scalar product like the volume form $\eps_{ijk}(\lambda)$ and the Hodge operator $*_\lambda$ which since now will be denoted by $\bar{\eps}_{ijk}$ and $\bar{*}$, respectively.          

Now let us calculate the derivative of $\xi^A(\lambda)$:
\begin{multline*}
\frac{d\xi^A}{d\lambda}=-\frac{1}{2}\veps^A{}_{BCD}\bar{\eps}^{ijk}{\theta}^B_i{\theta}^C_j\frac{d\theta^D_k}{d\lambda}=-\frac{1}{2}\veps^A{}_{BCD}\bar{\eps}^{ijk}{\theta}^B_i{\theta}^C_j(a_k\xi^D+\bar{\eps}^{ln}{}_kb_l\theta^D_n)=\\=-\frac{1}{2}\veps^A{}_{BCD}\bar{\eps}^{ijk}{\theta}^B_i{\theta}^C_ja_k\xi^D-\frac{1}{2}\veps^A{}_{BCD}(\bar{q}^{il}\bar{q}^{jn}-\bar{q}^{in}\bar{q}^{lj}){\theta}^B_i{\theta}^C_jb_l\theta^D_n=\\=-\frac{1}{2}\veps^A{}_{BCD}\bar{\eps}^{ijk}{\theta}^B_i{\theta}^C_ja_k\xi^D=\bar{*}\Big(-\frac{1}{2}\veps^A{}_{BCD}\theta^B(\lambda)\we\theta^C(\lambda)\xi^D(\lambda)\we a(\lambda)\Big).
\end{multline*} 
As shown in \cite{os}
\[
-\frac{1}{2}\veps^A{}_{BCD}\theta^B\we\theta^C\xi^D=*\theta^A,
\]
where $*$ is given by $\theta^A$. Since the Hodge operator $*_\lambda$ defined by $\theta^{A}(\lambda)$ coincides with $\bar{*}$ we have for all $\theta^A(\lambda)$
\[
-\frac{1}{2}\veps^A{}_{BCD}\theta^B(\lambda)\we\theta^C(\lambda)\xi^D(\lambda)=\bar{*}\theta^A(\lambda)
\]   
On the other hand for any one-form $\alpha$ and any $k$-form $\beta$ \cite{os}
\begin{equation}
*(*\beta\we\alpha)=\vec{\alpha}\lr\beta.
\label{a-b}
\end{equation}
Using these two results we obtain
\begin{equation}
\frac{d\xi^A}{d\lambda}=\bar{*}(\bar{*}\theta^A\we a)=\vec{a}\lr\theta^A=\bar{q}^{kj}a_k(\lambda)\theta^A_j(\lambda).
\label{dxi-dl}
\end{equation}

We see now that the derivative of $\theta^A_i$ in \eqref{t-trans} is a linear combination of $\xi^A$ and $\theta^A_k$ and the derivative of $\xi^A$ in \eqref{dxi-dl} is a linear combination of $\theta^A_j$. Let us then consider the following system of linear differential equations:
\begin{equation}
\frac{d\zeta^A_\alpha}{d\lambda}=K_\alpha{}^\beta(\lambda)\,\zeta^A_\beta, \ \ \ \ \   A,\alpha,\beta=0,1,2,3,
\label{zeta}
\end{equation}
where 
\[
\Big(K_\alpha{}^\beta(\lambda)\Big)=
\begin{pmatrix}
0 & K_0{}^j(\lambda)\\
K_i{}^0(\lambda) & K_i{}^j(\lambda)
\end{pmatrix}=
\begin{pmatrix}
0 & \bar{q}^{kj}a_k(\lambda)\\
a_i(\lambda) & \bar{\eps}^{kj}{}_ib_k(\lambda)
\end{pmatrix}.
\]
It is clear that a solutions $\zeta^A_\alpha(\lambda)$ of \eqref{zeta} with the initial condition
\begin{align*}
& \zeta^A_0(0)=\bar{\xi}^A, &&\zeta^A_i(0)=\bar{\theta}^A_i, 
\end{align*}
where 
\[
\bar{\xi}^A\equiv\xi^A(0)
\]
corresponds to $\bar{\theta}^A_i$, provides us with the solution $\theta^A_i(\lambda)$ of \eqref{t-trans} given by the initial condition \eqref{ini} together with the corresponding values of $\xi^A(\lambda)$:
\begin{align*}
&\xi^A(\lambda)=\zeta^A_0(\lambda),&&\theta^A_i(\lambda)=\zeta^A_i(\lambda).
\end{align*}

This particular solution $\zeta^A_\alpha(\lambda)$ defines a scalar product on $\R^4$ 
\begin{equation}
\Big(h_{\alpha\beta}(\lambda)\Big):=\Big(\eta_{AB}\zeta^A_\alpha(\lambda)\zeta^B_\beta(\lambda)\Big)=
\begin{pmatrix}
\eta_{AB}\xi^A(\lambda)\xi^B(\lambda) & \eta_{AB}\xi^A(\lambda)\theta^B_j(\lambda) \\
\eta_{AB}\theta^A_i(\lambda)\xi^B(\lambda) & \eta_{AB}\theta^A_i(\lambda)\theta^B_j(\lambda)
\end{pmatrix}=
\begin{pmatrix}
-1 &0 \\
0 &\bar{q}_{ij}
\end{pmatrix}
\label{h-metr},
\end{equation}
which actually does not depend on $\lambda$. On the other hand the general solution of \eqref{zeta} reads  
\begin{equation}
\zeta^A_\alpha(\lambda)=\Lambda_\alpha{}^\beta(\lambda)\zeta^A_\beta(0)
\label{lor}
\end{equation}
with the matrix $(\Lambda_\alpha{}^\beta(\lambda))$ independent of the choice of the initial values $\zeta^A_\beta(0)$. Setting this to \eqref{h-metr} we obtain
\[
h_{\alpha\beta}=\eta_{AB}\Lambda_\alpha{}^\gamma(\lambda)\zeta^A_\gamma(0)\Lambda_\beta{}^\delta(\lambda)\zeta^B_\delta(0)=\Lambda_\alpha{}^\gamma(\lambda)\Lambda_\beta{}^\delta(\lambda)h_{\gamma\delta},
\]
which means that $(\Lambda_\alpha{}^\beta(\lambda))$ preserves the Lorentzian scalar product \eqref{h-metr}. Thus the matrix $(\Lambda_\alpha{}^\beta(\lambda))$ is an element of the Lorentz group in a non-standard (unless $\bar{q}_{ij}=\delta_{ij}$) representation.

Note now that we can choose a basis $(dx^i)$ of $T^*_x\Sigma$ in such a way that $\bar{q}_{ij}=\delta_{ij}$. Then $(\Lambda_\alpha{}^\beta(\lambda))$ is a matrix of the standard representation of the Lorentz group. Similarly, $(K_\alpha{}^\beta)$ is then a matrix of the Lie algebra of the Lorentz group in its standard representation:        
\[
\Big(K_\alpha{}^\beta\Big)=
\begin{pmatrix}
0 & a_1 &a_2&a_3\\
a_1& 0 & -b_3 & b_2\\
a_2& b_3 & 0 &-b_1\\
a_3& -b_2 & b_1 & 0 
\end{pmatrix}.
\]

The conclusion is that the gauge transformations of $\bar{\theta}^A_i$ generated by $B(a)$ and $R(b)$  correspond to the Lorentz transformations \eqref{lor} which preserve the scalar product \eqref{h-metr} and act on the tetrad $(\bar{\xi}^A,\bar{\theta}^A_i)=(\zeta^A_0(0),\zeta^A_i(0))$ as follows: 
\begin{equation}
\begin{pmatrix}
\bar{\xi}^A\\
\bar{\theta}^A_i
\end{pmatrix}\mapsto
\begin{pmatrix}
\Lambda_0{}^0(\lambda)&\Lambda_0{}^j(\lambda)\\
\Lambda_i{}^0(\lambda)&\Lambda_i{}^j(\lambda)
\end{pmatrix}
\begin{pmatrix}
\bar{\xi}^A\\
\bar{\theta}^A_j
\end{pmatrix}.
\label{nst-L-tr}
\end{equation}

The gauge transformations generated by $B(a)$ and $R(b)$ preserve the spacetime metric $g$. Indeed, according to \eqref{g-dec} $g$ is a function of the lapse $N$, the shift vector field $\vec{N}$ and the metric $q$. On the other hand, the transformations do not act on the lapse and the shift vector field and preserve the metric $q$.     
 
We mentioned in the introduction that the local Lorentz transformations generated by $B(a)$ and $R(b)$  act on the canonical variables in a non-standard way. Let us now clarify this statement. Since $(\theta^A)$ can be treated as a one-form on $\Sigma$ valued in the vector space $\mathbb{M}$ equipped with the Lorentzian scalar product $\eta$ it is natural to define local Lorentz transformations of $(\theta^A)$ as follows:
\begin{equation}
\theta^A\mapsto \Lambda^A{}_B\theta^B,
\label{st-L-tr}
\end{equation}
where $(\Lambda^A{}_B)$ is a field on $\Sigma$ valued in the group of linear isomorphism of $\mathbb{M}$ preserving the scalar product $\eta$, that is, valued in the Lorentz group. Comparing the formula above with \eqref{nst-L-tr} we see that the local Lorentz transformations generated by $B(a)$ and $R(b)$ act in a very different way than the standard transformations \eqref{st-L-tr}: the former ones act on the spatial index $i$ and mix components $\theta^A_i$ and $\xi^A$ of fixed $A$ while the latter ones act on the index $B$ related to a basis of $\mathbb{M}$ and mix components $\theta^B_i$ of fixed $i$.

\subsubsection{Hamiltonian formulation of TEGR versus a simple model described in \cite{os}}

The action \eqref{act} can be alternatively expressed as \cite{waldyr}
\begin{equation}
S[\bth^A]=-\frac{1}{2}\int\bed\bth^A\we\star\bed\bth_A-(\star \bed\star \bth^A)\we \bed\star\bth_A-\frac{1}{2}(\bed\bth^A\we\bth_A)\we\star(\bed\bth^B\we\bth_B).
\label{act-alt}
\end{equation}
Omitting the last two terms at the r.h.s. of this expression we obtain an action
\begin{equation}
s[\bth^A]=-\frac{1}{2}\int\bed\bth^A\we\star\bed\bth_A
\label{act-simple}
\end{equation}
defining the dynamics of a theory called Yang-Mills-type Teleparallel Model (YMTM) \cite{itin} canonical framework of which was studied in \cite{os}. The phase space of that theory coincides with the phase space of TEGR described in this paper. The Legendre transformation defined by \eqref{act-simple} turns out to be invertible (there are no primary constraints) and one obtains the following Hamiltonian: 
\begin{equation}
h[{\theta}^A,p_A,N,\vec{N}]=s(N)+v(\vec{N}),
\label{ham-simpl}
\end{equation}
where 
\[
s(N)=\int_\Sigma N\Big(\frac{1}{2}p^A\we{*}p_A-\xi^Adp_A+\frac{1}{2}{d}{\theta}^A\we{*}{d}{\theta}_A\Big)
\]
is a smeared scalar constraint and {$v(\vec{M})\equiv V(\vec{M})$} is a smeared vector constraint. These secondary constraints are the only constraints and they are of the first class:
\begin{align*}
&\{s(M),s({M'})\}=v(\vec{m}),\\ 
&\{v({\vec{M}}),s(M)\}=s({\cal L}_{\vec{M}}M),\\
&\{v(\vec{M}),v(\vec{M'})\}=v([\vec{M},\vec{M'}]),
\end{align*}
where $m$ is given by \eqref{m}. This means, in particular, that in this model there are no gauge transformations which could be interpreted as local Lorentz transformations. 

Taking YMTM as a reference point we see that the last two terms at the r.h.s. of \eqref{act-alt}  are responsible for the following features of this formulation of TEGR:
\begin{enumerate}
\item the non-invertibility of the Legendre transformation \eqref{leg-tr-0} hence
\item the presence of the primary constraints $B(a)$ and $R(b)$ hence 
\item the existence of gauge transformations interpreted as local Lorentz transformations; 
\item the more complicated form of the scalar constraint $S(N)$ hence  
\item the more complicated form of the Poisson bracket of the scalar constraints.
\end{enumerate}

\subsubsection{Structure of the scalar constraints \label{str-sc}}

Let us comment on the structure of the scalar constraints $S(M)$ of TEGR and $s(M)$ of YMTM\footnote{Description of the properties of $s(M)$ presented below comes form \cite{os}.}. Let $\bld{\alpha}=\bld{\alpha}^A\ot v_A$ and $\bld{\beta}=\bld{\beta}^B\ot v_B$ be two-forms on $\cal M$ valued in $\mathbb{M}$. Given cotetrad $(\bth^A)$ on the manifold, which via the metric $g$ defines the Hodge operator $\star$, one can introduce bilinear maps
\begin{align*}
(\bld{\alpha},\bld{\beta})\mapsto \bld{K}(\bld{\alpha},\bld{\beta})&:=\frac{1}{2}(\bld{\alpha}^A\we\bth_B)\we\star(\bld{\beta}^B\we\bth_A)-\frac{1}{4}(\bld{\alpha}^A\we\bth_A)\we\star(\bld{\beta}^B\we\bth_B),\\
(\bld{\alpha},\bld{\beta})\mapsto \bld{k}(\bld{\alpha},\bld{\beta})&:=\frac{1}{2}\bld{\alpha}^A\we\star\bld{\beta}_A
\end{align*}
valued in four-forms on $\cal M$. Similarly, let ${\alpha}={\alpha}^A\ot v_A$ and ${\beta}={\beta}^B\ot v_B$ be two-forms on $\Sigma$ valued in $\mathbb{M}$. Given restricted cotetrad $(\theta^A\equiv\un{\bth}^A)$ on the manifold, which via the metric $q$ defines the Hodge operator $*$, one can introduce bilinear maps
\begin{align*}
({\alpha},{\beta})\mapsto {K}({\alpha},{\beta})&:=\frac{1}{2}({\alpha}^A\we\theta_B)\we*({\beta}^B\we\theta_A)-\frac{1}{4}({\alpha}^A\we\theta_A)\we*({\beta}^B\we\theta_B),\\
({\alpha},{\beta})\mapsto {k}({\alpha},{\beta})&:=\frac{1}{2}{\alpha}^A\we*{\beta}_A
\end{align*}
valued in three-forms on $\Sigma$. Note now that the actions \eqref{act} of TEGR and \eqref{act-simple} of YMTM can be written respectively as
\begin{align*}
S[\bth^A]&=-\int \bld{K}(\bed \bth,\bed \bth), &s[\bth^A]&=-\int \bld{k}(\bed \bth,\bed \bth),
\end{align*}
where $\bed \bth=\bed \bth^A\ot v_A$. On the other hand the scalar constraints $S(M)$ and $s(M)$ can be expressed as 
\begin{align*}
S(M)&=\int {K}(p,p)-\xi^A dp_A +{K}(d\theta,d\theta),& s(M)&=\int {k}(p,p)-\xi^A dp_A +{k}(d\theta,d\theta)
\end{align*}
where $d\theta=d\theta^B\ot v_B$. 

We see thus that the form of each scalar constraint is closely related to the form of the corresponding action. Moreover, the relations in both cases of TEGR and YMTM follow the same pattern.

\subsubsection{Comparison with the Hamiltonian formulation of TEGR presented in \cite{wall-av} \label{wall-sc}}

{The action \eqref{act} was earlier used by Wallner \cite{wall-av} to derive a Hamiltonian formulation of TEGR. Since our formulation is based on the same action and uses $3+1$ decomposition techniques patterned on those by Wallner \cite{ham-diff,wall-av} a detailed comparison of both formulations is needed to reveal differences between them.}

Let us begin with a brief description of the $3+1$ decomposition of differential forms on $\cal M$ applied in \cite{wall-av}. $3+1$ decomposition of fields on $\cal M$ requires a prior choice of a foliation of $\cal M$. Wallner chooses such a foliation to be local while in this paper we assume a global foliation, however this difference is not essential and therefore it will be neglected in further considerations.    

To define a foliation of $\cal M$ Wallner assumes that a Lorentzian metric $g$ on $\cal M$ is given and chooses a time-like one-form $n$ such that $\bed n\we n=0$. By the Frobenius theorem a distribution defined by annihilators of $n$ is integrable and provides a foliation of $\cal M$. Then there exists a function $t$ on $\cal M$ such that every leaf of the foliation is distinguished by a condition $t={\rm const.}$ and  $n=f\bed t$ for a function $f$ on $\cal M$. Without loss of generality Wallner sets $f=-N^2$ where $N$ is the lapse function defined by $g$ and $\bed t$: $g^{-1}(\bed t,\bed t)=-N^{-2}$, where $g^{-1}$ is a metric inverse to $g$.  Then he decomposes a $k$-form $\alpha$ as follows
\begin{equation}
\alpha=\,^{\perp\!^w}\!\alpha+\un{\alpha}_{_w},
\label{wall-dec}
\end{equation}
where 
\begin{align*}
\,^{\perp\!^w}\!\alpha&:=\bed t\we \alpha_{\perp\!^w},& \alpha_{\perp\!^w}&:=T\lr\,\alpha,& \un{\alpha}_{_w}&:=T\lr(\bed t\we\alpha).  
\end{align*}
In these formulae $T$ is a vector field on $\cal M$ defined by ``raising'' the index of $n$ by the inverse metric $g^{-1}$:
\begin{equation}
T:=g^{-1}(n,\cdot)
\label{T}
\end{equation}
Obviously, $T$ is orthogonal in the sense of $g$ to the foliation defined by $n$  and $Tt=1$.

Let us note that at this point there is no essential difference between the Wallner's decomposition \eqref{wall-dec} of forms and one applied in the present paper (see Section \ref{dec-forms}). Indeed, both decompositions are defined by a one-form $\bed t$ and a vector field ($T$ or $\partial_t$) such that the value of the one-form  on the vector field is $1$ and a particular method of introducing the one-form and the vector field is irrelevant for the decomposition---if fact, the one-form $\bed t$ and the vector field $\partial_t$ used in this paper can be seen as originating from a metric $g$ on $\cal M$ via the Wallner's construction outlined above.                  

However, an essential difference can be seen in $3+1$ decompositions of a cotetrad $(\bth^A)$. To reveal the difference let us consider the restricted configuration space $\bld{\Theta}$, the foliation $\{\Sigma_t\}_{t\in\R}$, the function $t$, the vector field $\partial_t$ and an adapted coordinate frame $(t,x^i)$ all these introduced in Section \ref{3+1-M}. Recall that every $(\bth^A)\in\bld{\Theta}$ defines via \eqref{g} a Lorentzian metric $g$ on $\cal M$. Its inverse metric $g^{-1}$ reads \cite{os} 
\begin{equation}
g^{-1}=\frac{1}{N^2}\Big(-\partial_t\ot\partial_t+N^i\partial_t\ot\partial_i+N^i\partial_i\ot\partial_t+(N^2q^{ij}-N^iN^j)\partial_i\ot\partial_j\Big),
\label{g-1-dec}
\end{equation}
where $N,\vec{N}$ are the lapse function and the shift vector field given by \eqref{theta-0-xi-a} and $q^{ij}$ are components of the metric \eqref{q-1}. It is clear that $\bed t$ is a time-like one-form with respect to $g$, hence the foliation $\{\Sigma_t\}_{t\in\R}$ can be thought as one defined by this $g$ and  a time-like one-form $n=-N^2\bed t$ according to the Wallner's prescription. Note now that to decompose elements of $\bld{\Theta}$ we can use the one-form $\bed t$ and either    
\begin{enumerate}
\item fix a vector field $\tilde{T}$ transversal to the foliation such that $\tilde{T}t=1$ and decompose every $(\bth^A)\in\bld{\Theta}$ using this fixed vector field, or    
\item given $(\bth^A)\in\bld{\Theta}$,  define a Lorentzian metric $g$ via \eqref{g} and $T$ via \eqref{T} with $n$ given by $g$ and $\bed t$ and then decompose $(\bth^A)$ using this $T$; in other words, we may decompose $(\bth^A)$ by means of the $(\bth^A)$-dependent vector field $T$.     
\end{enumerate}    

Clearly, in this paper we apply the first option with $\partial_t$ being the fixed vector field. Wallner neither fixes {\em explicitely} a vector field to decompose all cotetrad fields nor states {\em explicitely} that each cotetrad $(\bth^A)$ is decomposed with respect to the $(\bth^A)$-dependent $T$. Nevertheless, there is a way to arrive at a definite conclusion that he applies the second option. Note first that, given vector field transversal to the foliation $\{\Sigma_t\}_{t\in\R}$,  there are many cotetrads in $\bld{\Theta}$ which generate metrics ``incompatible'' with the vector field, where ``incompatibility'' of a metric means here that the vector field is not orthogonal in the sense of the metric to the foliation. It is easy to see that the formulae (A.19) (except the last one) and the identity (A.20) in \cite{wall-av} are not true if a metric defining the Hodge operator $\star$ (denoted in \cite{wall-av} by $*$) is incompatible with the vector field defining the decomposition. Moreover, in such a case a formula for the first fundamental form $h$ of $\Sigma_t$ induced by $g$:
\[
h=\eta_{AB}\,\un{\bth}^A_{_w}\ot\un{\bth}^B_{_w}
\]
presented in \cite{wall-av} between the expressions (A.20) and (A.21) is not true either. Consequently, to ensure validity of all these formulae one should either $(i)$ apply the first option together with a gauge fixing which excludes those cotetrads for which these formulae are not true or $(ii)$ apply the second option. Since Wallner derives his formulation without any gauge fixing\footnote{Let us note that a statement to be found at the very beginning of Section II in \cite{wall-av} that the $\R^4$-valued one-form ``denotes a one-form basis orthonormal with respect to a metric $g$'' should not be interpreted as a restriction (gauge fixing) imposed on cotetrads because $(i)$ a similar statement at the beginning of Page 4280 is followed by a remark that ``its orthonormality means no restriction to the geometry of $\cal M$'' and $(ii)$ a restriction to cotetrads defining a fixed metric $g$ would not be compatible with the goal of \cite{wall-av} being a reexamination of the Ashtekar's variables. Thus the statement means rather that the one-form basis {\em defines} a metric $g$ via \eqref{g}.} we conclude that he applies the second option.   

A conclusion is that we decompose cotetrad fields in a different way than Wallner. Let us then compare both decompositions. By virtue of \eqref{g-1-dec} (see also \cite{mielke})
\[
T=\partial_t-\vec{N},
\] 
hence
\begin{equation}
\bth^A_{\perp\!^w}=\bth^A_\perp-\vec{N}\lr\bth^A=\bth^A_\perp-\vec{N}\lr\un{\bth}^A=N\xi^A
\label{bth_perp_w}
\end{equation}
(where we used \eqref{theta-0-xi-a} in the last step) and
\begin{equation}
\un{\bth}^A_{_w}=T\lr(\bed t\we\bth^A)=\un{\bth}^A+\bed t \vec{N}\lr\bth^A=\un{\bth}^A+\bed t \vec{N}\lr\un{\bth}^A.
\label{bth_un_w}
\end{equation}

Now we are able to list some important differences between both Hamiltonian formulations:\begin{enumerate}
\item the Wallner's ``position'' variable $(\un{\bth}^A_{_w})$ does not coincide with our $(\un{\bth}^A)\equiv(\theta^A)$; 
\item in the Wallner's formulation the four functions $(\bth^A_{\perp\!^w})=(N\xi^A)$ are non-dynamical variables, while here the non-dynamical variables are the lapse $N$ and the shift $\vec{N}$; 
\item Wallner introduces momenta conjugate to his non-dynamical variables $(\bth^A_{\perp\!^w})$ (the momenta are obviously constrained to be zero), while here we treat the non-dynamical variables as Lagrangian multipliers; consequently, Wallner works with the full phase space while we work with a reduced phase space\footnote{On Page 4268 Wallner mentions a possibility of reducing the phase space but the reduction is not carried out---see the description of the next difference.}; 
\item in \cite{wall-av} constraints of TEGR and a Hamiltonian (2.17) are not expressed as {\em explicite} functions of canonically conjugate variables---the time derivatives of $({\un{\bth}}^A_{_w})$ are not eradicated from a scalar constraint defined as the longitudinal part of (2.9)\footnote{In the first formula in (2.13) in \cite{wall-av} describing the scalar constraint the time derivatives of $({\un{\bth}}^A_{_w})$ appear explicitely. In further formulae (2.16a), (2.30) and (2.63) the time derivatives appear implicitly via  variables defined, respectively, by (2.11), (2.29) and (2.31)---it is clear from, respectively, (2.10b), (2.36) and (2.62) that these variables are not momenta conjugate to $({\un{\bth}}^A_{_w})$. On the other hand Wallner does not prove that these variables are functions on the phase space (since the Legendre transformation is not invertible not every function of the time derivatives of $({\un{\bth}}^A_{_w})$ is a function on the phase space).}; moreover in the constraints there appear the non-dynamical variables $(\bth^A_{\perp\!^w})$. In the present paper all constraints and the Hamiltonian are expressed explicitely in terms of the canonically conjugate variables $(\theta^A,p_B)$; moreover, the constraints do not contain the non-dynamical variables $N$ and $\vec{N}$.     
\item unlike here, in \cite{wall-av} there is no explicite description of Lorentz gauge transformations generated by primary constraints; moreover, a constraint generating an action of spatial diffeomorphisms is not explicitely isolated;  
\item unlike here, in \cite{wall-av} a constraint algebra is not presented.
\item the time derivatives in the Hamiltonian field equations (A.43) and (A.44) in \cite{wall-av} are in fact Lie derivatives with respect to the $(\bth^A)$-dependent vector field $T$ (see the last formula in (A.13)). Therefore it is not clear whether the Wallner's formalism is able to describe the evolution of $(\bth^A)$ with respect to a vector field which is not orthogonal to the foliation $\{\Sigma_t\}_{t\in\R}$ in the sense of the metric $g$ defined by this $(\bth^A)$. The present formalism describes the evolution of all cotetrads in $\bld{\Theta}$ with respect to the fixed vector fields $\partial_t$, but since it is fixed arbitrarily any other vector field transversal to the foliation may be fixed instead of $\partial_t$. Thus the present formalism is able to describe the evolution of $(\bth^A)$ with respect to any vector field transversal to the foliation.     
\end{enumerate} 

There is also another important difference between the two formulations. Let us recall that usually while deriving a Hamiltonian formulation of a field theory one not only decomposes fields with respect to a foliation of a spacetime but also identifies decomposed fields with time-dependent fields defined on a space (i.e. on a three-dimensional manifold representing a space). Clearly, such an identification requires to define a family of diffeomorphisms such that each of them maps the space onto a leaf of the foliation. In this paper the identification of decomposed forms $\alpha_\perp$ and $\un{\alpha}$ with time-dependent forms on the space $\Sigma$ is carried out naturally by means of pull-back given by the diffeomorphisms $\{\varphi_t\}_{t\in\R}$ (see Section \ref{dec-forms}). In the Wallner's paper the decomposed forms {\em are not} identified with time-dependent fields on a space---his canonical variables are {\em space-time fields}. Let us note that in the case of the Wallner's forms $\bth^A_{\perp\!^w}$ and $\un{\bth}^A_{_w}$ every identification with time-dependent fields on $\Sigma$ defined by pull-back gives an unsatisfactory result---by virtue of \eqref{bth_perp_w} and \eqref{bth_un_w}
\begin{align*}
\tilde{\varphi}_{t*}\bth^A_{\perp\!^w}&=N\xi^A,&\tilde{\varphi}_{t*}\un{\bth}^A_{_w}&=\tilde{\varphi}_{t*}\un{\bth}^A,
\end{align*}
where $\tilde{\varphi}_t:\Sigma\to\Sigma_t$ is any diffeomorphism. This means that the resulting forms on $\Sigma$ do not contain the function $\vec{N}\lr\bth^A$. Consequently, this identification is not injective (distinct cotetrads are mapped to the same fields on $\Sigma$) and results in a gauge fixing $\vec{N}=0$.    

Let us finally note that by virtue of \eqref{bth_perp_w}, \eqref{bth_un_w} and \eqref{xixi}
\begin{equation}
\eta_{AB}\bth^A_{\perp\!^w}\un{\bth}^B_{_w}=N\xi_B(\un{\bth}^B+\bed t\vec{N}\lr\un{\bth}^B)=0,
\label{thth0}
\end{equation}
which means that these variables are not completely independent---in fact, $(\bth^A_{\perp\!^w})$ contains only one degree of freedom (per point) independent of $(\un{\bth}^A_{_w})$. Indeed, if $(\vec{X}_i)$, $i=1,2,3$, is a (local) frame on $\Sigma_t$ then the functions $(\xi^A)$ on $\Sigma_t$ can be alternatively defined by the first equation of \eqref{xixi} and $\vec{X}_i\lr\un{\bth}^A_{_w}\xi_A=0$ (see \eqref{bth_un_w}). Taking into account \eqref{bth_perp_w} we conclude that the only degree of freedom in $(\bth^A_{\perp\!^w})$ independent of $(\un{\bth}^A_{_w})$ is the lapse function. However, it seems that Wallner overlooked \eqref{thth0}---he treats $(\bth^A_{\perp\!^w})$ as four independent variables and using them derives four constraints (2.9) in \cite{wall-av}. This of course causes a doubt whether the derivation of these constraints is correct.

To summarize the discussion above we conclude that the Hamiltonian formulation of TEGR presented in this paper is essentially different from that of Wallner. 

{Let us also note that the Wallner's formulation is rather not very well suited for the Dirac's procedure of canonical quantization---to deal with constraints at the quantum level it is highly desirable (if not necessary) to know {\em explicitely} $(i)$ the constraints as functions on the phase space expressed in terms of canonically conjugate variables, $(ii)$ gauge transformations generated by constraints and $(iii)$ a constraint algebra and all these are lacking in the Wallner's analysis. The Wallner's formulation does not seem to be well suited for a background independent quantization either since in this formulation the diffeomorphism invariance of TEGR is rather hidden, in particular, it is not shown how spatial diffeomorphisms act on the Wallner's variables which are still space-time fields. Moreover, well developed methods of background independent canonical quantization applied in LQG \cite{rev,rev-1} suggest that classical elementary variables for such a quantization should be associated with some submanifolds of a space (in LQG classical elementary variables  are cylindrical functions associated with graphs and fluxes of momentum variables through two-dimensional surfaces). It is easier to work with elementary variables of this sort if (unlike in \cite{wall-av}) canonical variables are fields on the space. }               

\subsubsection{Comparison with the Hamiltonian formulation of TEGR presented in \cite{maluf} \label{sec-comp}}

A complete analysis of a Hamiltonian framework of TEGR considered as a theory of cotetrad fields was presented in \cite{maluf}\footnote{More precisely, the authors of \cite{maluf} consider TEGR with the unimodular condition imposed but it is easy to read off from their results the Hamiltonian formulation of the standard TEGR.}. The main difference between the approach of \cite{maluf} and that presented in this paper consists in the different way of parameterizing the non-dynamical part of the configurational degrees of freedom: in \cite{maluf} it is parameterized naturally by $\bth^A_\perp$, here we use the lapse $N$ and the shift $\vec{N}$ (see Equation \eqref{theta-0-xi-a}). Moreover, in \cite{maluf} an other action than \eqref{act} was used as the starting point of the analysis. Consequently, the resulting Hamiltonian, the set of constraints and the constraint algebra differ significantly from those derived in this paper. Formulae describing the secondary constraints $C^{\prime a}$ in \cite{maluf} seem to be a bit more complicated than the corresponding formulae \eqref{sec-constr}. Moreover, it is difficult to find a similarity between the structure of the constraints $C^{\prime a}$ and the structure of the original action analogous to that described in {Section \ref{str-sc}}.  Nevertheless, the constraint algebra presented in \cite{maluf} is much simpler than that described here---it is in fact a true Lie algebra---and all the constraints are of the first class.
   
Let us note finally that the primary constraints here and those in \cite{maluf}  generate local Lorentz transformations of the canonical variables. However, the transformations in \cite{maluf} are the standard ones \eqref{st-L-tr} while here we obtained the non-standard transformations \eqref{nst-L-tr}. It is a bit surprising that such a seemingly innocent change in parameterization of the non-dynamical degrees of freedom results in an essential change of the action of local Lorentz transformations.     

\section{Derivation of the Hamiltonian}

Let us recall that to describe the canonical framework of TEGR we use a Hamiltonian formalism adapted to differential forms \cite{ham-diff,mielke} (see also \cite{os}). 

\subsection{$3+1$ decomposition of the action}

It was shown in \cite{os} that if $\alpha,\beta$ are $k$-forms on $\cal M$ and $\star$ is the Hodge operator given by the spacetime metric $g$ (defined by Equation \eqref{g}) then
\begin{equation}
\alpha\we\star\beta=-N^{-1}\bed t\we(\alpha_\perp-\vec{N}\lrcorner\un{\alpha})\we{*}(\beta_\perp-\vec{N}\lrcorner\un{\beta})+N\,\bed t\we\un{\alpha}\we{*}\,\un{\beta},
\label{A*B-dec}
\end{equation}
where ${*}$ is the Hodge operator given by the Riemannian metric $q$ (defined by Equation \eqref{q}) on $\Sigma_t$, and $N$ and $\vec{N}$ are, respectively, the lapse and the shift appearing in \eqref{theta-0-xi-a}\footnote{In fact, to prove \eqref{A*B-dec} it is not necessary to assume that the spacetime metric $g$ is defined by a cotetrad---it is sufficient to assume \eqref{g-dec} and \eqref{N>0}.}. 

To obtain a $3+1$ decomposition of the action \eqref{act} we apply the decomposition \eqref{A*B-dec} separately to the first and the second terms under the integral at the r.h.s. of \eqref{act}. By virtue of \eqref{perp-un}
\begin{equation*}
(\bed \bth^A\we\bth_B)_\perp=(\bed\bth^A)_\perp\we\un{\bth}_B+\un{\bed \bth}^A\we(\bth_B)_\perp=\LL_{\partial_t}\un{\bth}^A\we\un{\bth}_B-d(\bth^A_\perp)\we\un{\bth}_B+d\un{\bth}^A\we(\bth_B)_\perp.
\end{equation*}
and
\[
\un{\bed \bth^A\we\bth_B}=d\un{\bth}^A\we\un{\bth}_B.
\]
In order to make further calculations more transparent we introduce the following abbreviations:
\begin{gather}
F^A{}_B\equiv{d}\un{\bth}^A\we\un{\bth}_B,\label{F}\\
E^A{}_B\equiv-{d}(\bth^A_\perp)\we\un{\bth}_B+{d}\un{\bth}^A\we\bth_{B\perp}-\vec{N}\lrcorner F^A{}_B.
\label{E}
\end{gather}
Since now we will moreover apply the simplified notation \eqref{bt-t} and \eqref{Ltbt}. Now we can write
\[
(\bed \bth^A\we\bth_B)_\perp-\vec{N}\lr\un{\bed \bth^A\we\bth_B}=\dot{\theta}^A\we\theta_B+E^A{}_B.
\]
At this point we can easily decompose the action \eqref{act} obtaining thereby
\begin{multline}
S[\theta^A,N,\vec{N}]=\int \frac{1}{2N}\bld{d}t\we({\dot{\theta}}^A\we{\theta}_B+E^A{}_B )\we{*}({\dot{\theta}}^B\we{\theta}_A+E^B{}_A )-\frac{N}{2}\bld{d}t\we F^A{}_B\we{*}F^B{}_A-\\
-\frac{1}{4N}\bld{d}t\we({\dot{\theta}}^A\we{\theta}_A+E^A{}_A )\we{*}({\dot{\theta}}^B\we{\theta}_B+E^B{}_B )+\frac{N}{4}\bld{d}t\we F^A{}_A\we{*}F^B{}_B.
\label{act-dec}
\end{multline}

\subsection{Legendre transformation}

Note that in the decomposed action \eqref{act-dec} there is no Lie derivative of $N$ and $\vec{N}$ with respect to $\partial_t$. Therefore since now we will treat the lapse and the shift as Lagrange multipliers. Consequently, we are left with one-forms $(\theta^A)$ as the only configurational variables which are dynamical. Thus a point in the phase space of the theory is a collection $(\theta^A,p_B)$ $(A,B=0,1,2,3)$, where $(\theta^A)$ are one-forms on $\Sigma$ such that the metric \eqref{q-tt} is Riemannian and the momentum $p_A$ conjugate to $\theta^A$ is a two-form on $\Sigma$.

Let us recall that we denoted by $L$ the four-form on $\cal M$ being the integrand in \eqref{act}. The Legendre transformation reads
\begin{equation}
p_A=\frac{\partial L_\perp}{\partial{\dot{\theta}}^A}
=N^{-1}\Big({\theta}_B\we{*}({\dot{\theta}}^B\we{\theta}_A+E^B{}_A )-\frac{1}{2}{\theta}_A\we{*}({\dot{\theta}}^B\we{\theta}_B+E^B{}_B )\Big).
\label{leg-tr}
\end{equation}
and allows us to introduce a Hamiltonian
\begin{multline}
H_0[{\theta}^A,{\dot{\theta}}^A,N,\vec{N}]:=\int_\Sigma{\dot{\theta}}^A\we p_A-L_\perp=\int_\Sigma \frac{1}{2N}({\dot{\theta}}^A\we{\theta}_B-E^A{}_B )\we{*}({\dot{\theta}}^B\we{\theta}_A+E^B{}_A )-\\
-\frac{1}{4N}({\dot{\theta}}^A\we{\theta}_A-E^A{}_A )\we{*}({\dot{\theta}}^B\we{\theta}_B+E^B{}_B )+\frac{N}{2}F^A{}_B\we{*}F^B{}_A-\frac{N}{4}F^A{}_A\we{*}F^B{}_B.
\label{H0-conf}
\end{multline}
expressed as a functional depending on $\theta^A$, Lie derivatives $\dot{\theta}^A$, the lapse and the shift. In other words, this Hamiltonian is a functional on the restricted configuration space. Of course, our goal is to obtain a Hamiltonian defined on the Cartesian product of the phase space and a space of all  Lagrange multipliers, that is, lapse functions and shift vector fields. As a first attempt to reach the goal we will try to invert the Legendre transformation \eqref{leg-tr}. 

Let us start by acting on both sides of \eqref{leg-tr} by the Hodge operator ${*}$---using \eqref{a-b} we obtain
\begin{equation}
N{*}p_A+\vec{\theta}_B\lrcorner E^{B}{}_A-\frac{1}{2}\vec{\theta}_A\lrcorner E^B{}_B=-\vec{\theta}_B\lrcorner({\dot{\theta}}^B\we{\theta}_A)+\frac{1}{2}\vec{\theta}_A\lrcorner({\dot{\theta}}^B\we{\theta}_B)
\label{np-dott}
\end{equation}
Denoting
\begin{equation}
\pi_A\equiv N{*}p_A+\vec{\theta}_B\lrcorner E^{B}{}_A-\frac{1}{2}\vec{\theta}_A\lrcorner E^B{}_B
\label{pi-df}
\end{equation}
we rewrite the result above in the following form:
\begin{equation}
\pi_{Aj}=\theta_{B(i}\dot{\theta}^B_{j)}\theta^i_{A}-\theta^i_B\dot{\theta}^B_i\theta_{Aj}=\theta_{B(i}\dot{\theta}^B_{j)}\theta^i_{A}-\theta_{B(i}\dot{\theta}^B_{k)}q^{ik}\theta_{Aj}.
\label{pi}
\end{equation}
Note  that by virtue of \eqref{q}
\begin{equation}
\theta^A_i \theta^{j}_{A}=\theta^A_i \theta_{Ak}q^{kj}=q_{ik}q^{kj}=\delta^j{}_i.
\label{tt-delta}
\end{equation}
Using this identity we obtain from \eqref{pi} 
\[
\pi_{Aj}\theta^A_k=\theta_{B(j}\dot{\theta}^B_{k)}-\theta^i_B\dot{\theta}^B_iq_{jk}.
\]
Contracting both sides of the last formula with $q^{jk}$ we get
\[
\pi_{Aj}\theta^{Aj}=-2\theta^i_B\dot{\theta}^B_i.
\] 
Thus
\begin{equation}
\pi_{Aj}\theta^A_k-\frac{1}{2}\pi_{Ai}\theta^{Ai}q_{jk}=\theta_{A(j}\dot{\theta}^A_{k)}.
\label{pi-vel}
\end{equation}

It is evident now that the Legendre transformation is not invertible. The source of the non-invertibility is twofold:
\begin{enumerate}
\item treating $\dot{\theta}^A_i$ of a fixed $i$ as a four-component vector we see that in the expression \eqref{pi-vel} there appear only contractions of $\dot{\theta}^A_i$ with the three linearly independent vectors $\{\theta^A_j\}$ ($j=1,2,3$) while the contraction $\xi_A\dot{\theta}^A_i$ is missing (recall that at each point of $\Sigma$ the values of functions $(\xi^A,\theta^A_i)$ form a basis of $\mathbb{M}$).  
\item only the symmetric part of the tensor $\theta_{Ai}\dot{\theta}^A_{j}$ appears in the expression.
\end{enumerate}
This means that information encoded in $\dot{\theta}^A$ is reduced by the transformation. To analyze the reduction let us fix a point $x\in\Sigma$ and values of $\theta^A$, $d\theta^A$, $N$ and $\vec{N}$ at this point and treat \eqref{leg-tr} as a map  transforming $\dot{\theta}^A(x)$ to $p_A(x)$. This map can be seen as a composition $I\circ P$ of an injection $I$ and a linear projection $P$. Indeed, $P$ is a map which maps 12 independent quantities $\dot{\theta}^A_i(x)$ to $\theta_{A(j}(x)\dot{\theta}^A_{k)}(x)$ loosing information encoded in 6 quantities $\xi_A(x)\dot{\theta}^A_i(x)$ and $\theta_{A[j}(x)\dot{\theta}^A_{k]}(x)$. It follows from Equations \eqref{pi} and \eqref{pi-df} that the value $\theta_{A(j}(x)\dot{\theta}^A_{k)}(x)$ unambiguously gives the value $p^A_{ij}(x)$ and this mapping is what we called $I$ above. On the other hand we see from Equation \eqref{pi-vel} that once we have $p^A_{ij}(x)$ we have also $\theta_{A(j}(x)\dot{\theta}^A_{k)}(x)$ which means that $I$ is an injection. Hence the image of the map $I\circ P:\dot{\theta}^A(x)\mapsto p_A(x)$ is $6$-dimensional. Therefore there should be $6$ independent constraints imposed on 12 quantities $p_{Aij}(x)$: 
\begin{enumerate}
\item contracting both sides of \eqref{np-dott} with $\xi^A$ and taking into account \eqref{pi-df} and \eqref{xixi} we obtain
\begin{equation}
\pi_A\xi^A=0
\label{c-1}
\end{equation}
\item extracting the antisymmetric part of both sides of \eqref{pi-vel} we obtain the three remaining constraints
\begin{equation}
\pi_{A[j}\theta^A_{k]}=0
\label{c-anti-0}
\end{equation}
or equivalently
\begin{equation}
\pi_A\we{\theta}^A=0.
\label{c-anti}
\end{equation}
\end{enumerate}  
Note that the conditions \eqref{c-1} and \eqref{c-anti} contain the one-form $\pi_A$ which depends on the laps $N$ and the shift $\vec{N}$. Therefore at this point it is not obvious that the conditions define {\em constraints on the phase space}.

\subsection{Primary constraints}

The goal of this section is to remove the lapse $N$ and the shift $\vec{N}$ from the conditions \eqref{c-1} and \eqref{c-anti}. In other words we will show the conditions are in fact {\em primary constraints}. Moreover, we will prove there that they are no other primary constraint than those defined by \eqref{c-1} and \eqref{c-anti}. 

Let us start by stating and proving two auxiliary identities:
\begin{align}
\vec{\theta}^B\lrcorner{\theta}^A&=\eta^{AB}+\xi^A\xi^B, \label{tt-eta}\\
\theta^A\we(\vth_A\lr\alpha)&=k\alpha \label{tta=ka}
\end{align}
for any $k$-form $\alpha$ on $\Sigma$.
\begin{proof}[Proof of \eqref{tt-eta}]
Using the components of the metric $g^{-1}$ inverse to $g$ to raise the space-time indeces (here: lower case Greek letters) we obtain from \eqref{g} 
\[
\bth_{A\alpha}\bth^{A\beta}=\delta^\beta{}_\alpha,
\]
which means that
\[
\bth_{A\alpha}\bth^{B\alpha}=\delta^{B}{}_A.
\]
Raising the index $A$ we obtain  
\[
\bth^A_\alpha\bth^B_\beta g^{\alpha\beta}=\eta^{AB}.
\]
Setting to this equation the components of $g^{-1}$ expressed as \cite{adm}
\[
g^{00}=-N^{-2}, \ \ g^{0i}=N^{-2}N^i, \ \ g^{ij}=q^{ij}-N^{-2}N^iN^j
\]
and taking into account that $\bth^A_0=\bth^A_\perp$ we obtain
\[
\eta^{AB}=-\frac{1}{N^2}(\bth^A_\perp-\vec{N}\lrcorner{\theta}^A)(\bth^B_\perp-\vec{N}\lrcorner{\theta}^B)+\vec{\theta}^B\lrcorner{\theta}^A=-\xi^A\xi^B+\vec{\theta}^B\lrcorner{\theta}^A,
\]
where in the last step we applied \eqref{theta-0-xi-a}.
\end{proof}
\begin{proof}[Proof of \eqref{tta=ka}]
Using \eqref{tt-delta} we calculate
\begin{multline*}
\theta^A\we(\vth_A\lr\alpha)=\theta^A_i dx^i\we (\theta^j_{A}\partial_j\lr\frac{1}{k!}\alpha_{a_1\ldots a_k}dx^{a_1}\we\ldots\we dx^{a_k})=\\=\theta^A_i \theta^{j}_{A}  \frac{k}{k!}\alpha_{ja_2\ldots a_k}dx^i\we dx^{a_2}\we\ldots\we dx^{a_k}=k\frac{1}{k!}\alpha_{ia_2\ldots a_k}dx^{i}\we dx^{a_2}\we\ldots\we dx^{a_k}=k\alpha.
\end{multline*}
\end{proof}

It will be convenient to denote 
\begin{equation}
\rho_A\equiv N{*}p_A+\vec{\theta}_B\lrcorner E^B{}_A.
\label{rho}
\end{equation}
Then 
\begin{equation}
\pi_A=\rho_A-\frac{1}{2}\vec{\theta}_A\lrcorner E^B{}_B.
\label{pi-rho}
\end{equation}
Now let us express all the forms above as explicite functions of ${\theta}^A,p_B,N,\vec{N}$. To this end we set into \eqref{E} the function $\bth^A_\perp$ written as in \eqref{theta-0-xi-a}. Then with application of \eqref{xixi}, \eqref{dxixi} and \eqref{tt-eta} we obtain in turn  
\begin{equation}
\begin{aligned}
&E^A{}_B=-\xi^A{d}N\we{\theta}_B-N({d}\xi^A\we{\theta}_B-{d}{\theta}^A\we\xi_B)-{\cal L}_{\vec{N}}{\theta}^A\we{\theta}_B,\\
&\vec{\theta}_A\lrcorner E^A{}_B=N\Big(-(\vec{\theta}_A\lrcorner{d}\xi^A){\theta}_B+{d}\xi_B+\vec{\theta}_A\lrcorner{d}{\theta}^A\xi_B\Big)-(\vec{\theta}_A\lrcorner{\cal L}_{\vec{N}}{\theta}^A){\theta}_B+{\cal L}_{\vec{N}}{\theta}^A\vec{\theta}_A\lrcorner{\theta}_B,\\
&E^A{}_A=-N({d}\xi^A\we{\theta}_A-{d}{\theta}^A\we\xi_A)-{\cal L}_{\vec{N}}{\theta}^A\we{\theta}_A=2N\xi^A d{\theta}_A-{\cal L}_{\vec{N}}{\theta}^A\we{\theta}_A,\\
&\rho_A=N\Big({*}p_A-(\vec{\theta}_C\lrcorner{d}\xi^C){\theta}_A+{d}\xi_A+\vec{\theta}_C\lrcorner{d}{\theta}^C\xi_A\Big)-(\vec{\theta}_C\lrcorner{\cal L}_{\vec{N}}{\theta}^C){\theta}_A+{\cal L}_{\vec{N}}{\theta}^C\vec{\theta}_C\lrcorner{\theta}_A,
\end{aligned}
\label{E-rho}
\end{equation}
where ${\cal L}_{\vec{N}}$ denotes the Lie derivative on $\Sigma$ with respect to the shift $\vec{N}$ (recall that $\LL_{\vec{N}}=d\circ\vec{N}\lr+\vec{N}\lr\circ d$).

The condition \eqref{c-1} can be simplified as follows
\[
0=\pi_A\xi^A=\rho_A\xi^A=N(\xi^A*p_A-\vth_A\lr d\theta^A)=N*(\xi^Ap_A+\theta^A\we *d\theta_A),
\]
where we have used \eqref{a-b}, \eqref{xixi} and \eqref{dxixi}. Consequently, taking into account \eqref{N>0} we get 
\begin{equation}
\theta^A\we *d\theta_A+\xi^Ap_A=0,
\label{C-1}
\end{equation}
which coincides with \eqref{B-constr}. On the other hand using \eqref{tt-eta}, \eqref{tta=ka}, \eqref{xixi} and \eqref{dxixi} we can transform \eqref{c-anti} as follows
\begin{multline*}
0=\theta^A\we\pi_A=\theta^A\we\rho_A-E^A{}_A=N(\theta^A\we*p_A+\theta^A\we d\xi_A)+\theta^A\we(\LL_{\vec{N}}\theta^C)(\eta_{CA}+\xi_C\xi_A)-\\-E^A{}_A=N(\theta^A\we*p_A+\theta^A\we d\xi_A)+\theta^A\we\LL_{\vec{N}}\theta_A-E^A{}_A=N(\theta^A\we*p_A-\xi_Ad\theta^A),
\end{multline*}
hence by virtue of \eqref{N>0}
\begin{equation}
\theta^A\we*p_A-\xi_Ad\theta^A=0
\label{C-anti}
\end{equation}
which coincides with \eqref{R-constr}. 

Let us fix a point $x\in\Sigma$ and values of $\theta^A$ and $d\theta^A$ at $x$. Then 12 quantities $p_{Aij}(x)$ can be fully encoded in 12 {\em independent} quantities 
\[
\xi^A(x)(*p_A)_i(x), \ \ \ \theta^A_{[j}(x)(*p_A)_{i]}(x), \ \ \ \theta^A_{(j}(x)(*p_A)_{i)}(x).
\]
Note now that the conditions \eqref{C-1} and \eqref{C-anti} fix values of the former two quantities (to see this act by $*$ on both sides of \eqref{C-1}). This means that these conditions are {\em independent}. Since there are 6 of them and since they do not contain the lapse and the shift they are 6 independent {\em primary constraints} on the phase space. 

Recall that in the previous subsection we concluded that, given values of $\theta^A$, $d\theta^A$, the lapse and the shift at $x$, there are 6 independent constraints imposed on $p_{Aij}(x)$ being values of the Legendre transformation \eqref{leg-tr}. This means that there are no other primary constraints than \eqref{C-1} and \eqref{C-anti}.        

Let us finally note that setting to \eqref{leg-tr} the two-forms $E^A{}_B$ and $E^B{}_B$ expressed as in \eqref{E-rho} we obtain the formula \eqref{p_A-exp}.

\subsection{The Hamiltonian $H_0$ as a functional of the canonical variables}

The non-invertibility of the Legendre transformation means that the Hamiltonian $H_0$ \eqref{H0-conf} can be defined only on a part of the phase space being the image of the transformation, that is, on a part distinguished by vanishing of the primary constraints \eqref{C-1} and \eqref{C-anti}. To replace in $H_0$ the ``velocities'' $\dot{\theta}^A$ by the momenta $p_A$ let us first show that the Hamiltonian depends on the ``velocities'' merely via the combination $\theta_{A(i}\dot{\theta}^A_{j)}$. 

Let us start by gathering the terms containing ${\dot{\theta}}^A$ in \eqref{H0-conf}:
\begin{multline}
H_0=\int_\Sigma \frac{1}{2N}\Big(({\dot{\theta}}^A\we{\theta}_B)\we{*}({\dot{\theta}}^B\we{\theta}_A)-\frac{1}{2}({\dot{\theta}}^A\we{\theta}_A)\we{*}({\dot{\theta}}^B\we{\theta}_B)\Big)-\\
-\frac{1}{2N} E^A{}_B\we{*}E^B{}_A 
+\frac{1}{4N}E^A{}_A \we{*}E^B{}_B+\frac{N}{2}F^A{}_B\we{*}F^B{}_A-\frac{N}{4}F^A{}_A\we{*}F^B{}_B.
\label{ham-vel}
\end{multline}
Consider now the following map acting on one-forms $\alpha^A,\beta^B$ on $\Sigma$:
\begin{equation}
(\alpha^A,\beta^B)\mapsto G(\alpha^A,\beta^B):=\Big((\alpha^A\we{\theta}_B)\we{*}(\beta^B\we{\theta}_A)-\frac{1}{2}(\alpha^A\we{\theta}_A)\we{*}(\beta^B\we{\theta}_B)\Big).
\label{GAB-0}
\end{equation}
This map can be used to rewrite the first two terms in \eqref{ham-vel} as $G(\dot{\theta}^A,\dot{\theta}^B)/2N$. Now let us express $G(\alpha^A,\beta^B)$ in terms of the components of the one-forms. Given $k$-forms $\gamma$ and $\gamma'$,  
\begin{align}
\gamma\we*\gamma'&=\scal{\gamma}{\gamma'}\eps, &\scal{\gamma}{\gamma'}&:=\frac{1}{k!}\gamma_{i_1\ldots i_k}\gamma'_{j_1\ldots j_k}q^{i_1j_1}\ldots q^{i_kj_k}.
\label{g*g}
\end{align}
If $\alpha$ is a one-form then $(\alpha\we\theta_B)_{ij}=\alpha_i\theta_{Bj}-\alpha_j\theta_{Bi}$. Therefore for one-forms $\alpha,\beta$    
\[
(\alpha\we{\theta}_B)\we{*}(\beta\we{\theta}_A)=\scal{\alpha\we{\theta}_B}{\beta\we{\theta}_A}\eps=(\scal{\alpha}{\beta}\scal{\theta_B}{\theta_A}-\scal{\alpha}{\theta_A}\scal{\beta}{\theta_B})\eps.
\]
From this result we can easily obtain formulae describing the first and the second term at the r.h.s. of \eqref{GAB-0}: $(i)$ to get the first one we set $\alpha=\alpha^A$ and $\beta=\beta^B$ and assume summation over $A$ and $B$, $(ii)$ to get the second one we exchange $\theta_B\leftrightarrow \theta_A$, set $\alpha=\alpha^A$ and $\beta=\beta^B$ and assume summation over $A$ and $B$. Thus
\[
G(\alpha^A,\beta^B)=\Big(\frac{1}{2}\big(\scal{\alpha^A}{\beta^B}\scal{\theta_B}{\theta_A}+\scal{\alpha^A}{\theta_B}\scal{\beta^B}{\theta_A}\big)-\scal{\alpha^A}{\theta_A}\scal{\beta^B}{\theta_B}\Big)\eps.
\]
The first two terms at the r.h.s. of the equation above can be rewritten as
\begin{multline*}
\frac{1}{2}\big(\theta_{Ai}\alpha^A_j \theta_{Bk}\beta^B_l q^{ik}q^{jl}+\theta_{Aj}\alpha^A_i \theta_{Bk}\beta^B_l q^{ik}q^{jl}\big)=\theta_{A(i}\alpha^A_{j)} \theta_{Bk}\beta^B_l q^{ik}q^{jl}=\\=\theta_{A(i}\alpha^A_{j)} \theta_{B(k}\beta^B_{l)} q^{ik}q^{jl}+\theta_{A(i}\alpha^A_{j)} \theta_{B[k}\beta^B_{l]} q^{ik}q^{jl}=\theta_{A(i}\alpha^A_{j)} \theta_{B(k}\beta^B_{l)} q^{ik}q^{jl},
\end{multline*} 
where the last step is based on the following fact:
\[
q^{ik}q^{jl}S_{ij}A_{kl}=0
\]        
if only $S_{ij}=S_{ji}$ and $A_{ij}=-A_{ji}$.  
Finally,
\begin{equation}
G(\alpha^A,\beta^B)=(q^{ik}q^{jl}-q^{ij}q^{kl})\theta_{A(i}\alpha^A_{j)}\theta_{B(k}\beta^B_{l)}\,{\eps}
\label{GAB}
\end{equation}
and consequently
\begin{multline}
H_0=\int_\Sigma \frac{1}{2N}\Big(G({\dot{\theta}}^A,{\dot{\theta}}^B)
- E^A{}_B\we{*}E^B{}_A +\frac{1}{2}E^A{}_A \we{*}E^B{}_B\Big)+\\
+\frac{N}{2}F^A{}_B\we{*}F^B{}_A-\frac{N}{4}F^A{}_A\we{*}F^B{}_B
\label{ham-vel-1}
\end{multline}
with
\begin{equation}
G({\dot{\theta}}^A,{\dot{\theta}}^B)=(q^{ik}q^{jl}-q^{ij}q^{kl})\theta_{A(i}\dot{\theta}^A_{j)}\theta_{B(k}\dot{\theta}^B_{l)}\,{\eps}.
\label{vel^2}
\end{equation}

Note now that by virtue of  Equations \eqref{pi-vel} and \eqref{q} 
\[
\theta_{A(j}\dot{\theta}^A_{k)}=\theta_{A(j}[\pi^A_{k)}-\frac{1}{2}(\vec{\theta}^C\lrcorner\pi_C)\theta^A_{k)}],
\]
provided $\pi_A$ satisfies \eqref{c-anti-0} (which is obviously satisfied by $(\theta^A,p_B)$ belonging to the image of the Legendre transformation). Thus the term \eqref{vel^2} can be expressed in the following form
\begin{multline*}
G(\dot{\theta}^A,{\dot{\theta}}^B)=(q^{ik}q^{jl}-q^{ij}q^{kl})\theta_{A(i}\dot{\theta}^A_{j)}\theta_{B(k}\dot{\theta}^B_{l)}\,{\eps}=\\=
(q^{ik}q^{jl}-q^{ij}q^{kl})\Big(\theta_{A(i}[\pi^A_{j)}-\frac{1}{2}(\vec{\theta}^C\lrcorner\pi_C)\theta^A_{j)}]\Big)\,\Big(\theta_{B(j}[\pi^B_{k)}-\frac{1}{2}(\vec{\theta}^D\lrcorner\pi_D)\theta^B_{k)}]\Big){\eps}=\\=G(\pi^A-\frac{1}{2}(\vec{\theta}^C\lrcorner\pi_C){\theta}^A,\pi^B-\frac{1}{2}(\vec{\theta}^D\lrcorner\pi_D){\theta}^B)
\end{multline*}
which allows us to rewrite \eqref{ham-vel-1} as
\begin{multline}
H_0[{\theta}^A,p_A,N,\vec{N}]=\int_\Sigma \frac{1}{2N}\Big[G\Big(\pi^A-\frac{1}{2}(\vec{\theta}^C\lrcorner\pi_C){\theta}^A,\pi^B-\frac{1}{2}(\vec{\theta}^D\lrcorner\pi_D){\theta}^B\Big)-\\
- E^A{}_B\we{*}E^B{}_A +\frac{1}{2}E^A{}_A \we{*}E^B{}_B\Big]+\frac{N}{2}F^A{}_B\we{*}F^B{}_A-\frac{N}{4}F^A{}_A\we{*}F^B{}_B,
\label{ham-pi-0}
\end{multline}
where $\pi_A$ is a function of ${\theta}^A,p_B,N,\vec{N}$ given by \eqref{pi-df}.

\subsection{An explicite form of the Hamiltonian $H_0$}

Now we begin quite a long series of transformations of the Hamiltonian \eqref{ham-pi-0} aimed at expressing it explicitely as a functional of the canonical variables ${\theta}^A$ and $p_A$, the lapse $N$ and the shift $\vec{N}$. Let us start by introducing and proving some auxiliary formulae which will be repeatedly used in the sequel.

\subsubsection{Auxiliary formulae \label{aux}}

For any one-forms $\alpha$ and $\beta$ the following formulae hold:  
\begin{align}
\alpha\we*\beta&=(\vec{\alpha}\lr\beta )\eps,\label{a-b-e}\\
(\alpha\we{\theta}^B)\we{*}(\beta\we{\theta}^A)&=-(\vec{\theta}^A\lrcorner\alpha)(\vec{\theta}^B\lrcorner\beta){\eps}+(\eta^{AB}+\xi^A\xi^B)\alpha\we{*}\beta.\label{atbt-0}
\end{align}
\begin{proof}[Proof of \eqref{a-b-e}]
This formula follows immediately from \eqref{g*g} since for every one-forms $\scal{\alpha}{\beta}=\vec{\alpha}\lr\beta$.
\end{proof}
\begin{proof}[Proof of \eqref{atbt-0}] Since $q$ is Riemannian the square of the Hodge operator is an identity: $**={\rm id}$. Therefore  
\begin{multline*}
(\alpha\we{\theta}^B)\we{*}(\beta\we{\theta}^A)=\alpha\we{*}{*}\Big({\theta}^B\we{*}(\beta\we{\theta}^A)\Big)=-\alpha\we{*}[\vec{\theta}^B\lrcorner(\beta\we{\theta}^A)]=\\=-\alpha\we{*}[(\vec{\theta}^B\lrcorner\beta){\theta}^A]+\alpha\we{*}(\beta\,\vec{\theta}^B\lrcorner{\theta}^A)=-(\vec{\theta}^A\lrcorner\alpha)(\vec{\theta}^B\lrcorner\beta){\eps}+(\eta^{AB}+\xi^A\xi^B)\alpha\we{*}\beta
\end{multline*} 
---here in the second step we used \eqref{a-b}, and in the last one \eqref{a-b-e} and \eqref{tt-eta}.
\end{proof}

It follows from \eqref{a-b-e} and \eqref{tt-delta} that 
\begin{equation}
\theta^A\we*\theta_A=(\vth^A\lr\theta_A)\eps=3\eps.
\label{tt-3e}
\end{equation}
Setting in \eqref{atbt-0} $\alpha=\alpha_A$ and $\beta=\beta_B$ and assuming summations over $A$ and $B$ we get
\begin{equation}
(\alpha_A\we{\theta}^B)\we{*}(\beta_B\we{\theta}^A)=-(\vec{\theta}^A\lrcorner\alpha_A)(\vec{\theta}^B\lrcorner\beta_B){\eps}+\alpha_A\we*\beta^A+(\xi^A\alpha_A)\we{*}(\xi^B\beta_B).
\label{atbt}
\end{equation}    
Similarly, setting in \eqref{atbt-0} $\alpha=\beta_B$ and $\beta=\alpha_A$ and assuming summations over $A$ and $B$ we obtain
\begin{equation}
(\beta_B\we{\theta}^B)\we{*}(\alpha_A\we{\theta}^A)=-(\vec{\theta}^A\lrcorner\beta_B)(\vec{\theta}^B\lrcorner\alpha_A){\eps}+\beta_A\we*\alpha^A+(\xi^B\beta_B)\we{*}(\xi^A\alpha_A).
\label{atbt-1}
\end{equation}    
Setting $\alpha_A={\theta}_A$ in \eqref{atbt} gives
\begin{equation}
({\theta}_A\we{\theta}^B)\we{*}(\beta_B\we{\theta}^A)=-2(\vec{\theta}^B\lrcorner\beta_B){\eps}=-2\beta_A\we{*}{\theta}^A
\label{tt-bt}
\end{equation}
---these equalities hold due to \eqref{tt-3e}, \eqref{a-b-e} and \eqref{xixi}. Assume that in the formula just obtained $\beta_B={\theta}_B$. Applying \eqref{tt-3e} we obtain
\begin{equation}
({\theta}_A\we{\theta}^B)\we{*}({\theta}_B\we{\theta}^A)=-6{\eps}.
\label{4t}
\end{equation}

\subsubsection{Calculations}

Since now till the end of the paper we will so often apply the formulae \eqref{xixi} and \eqref{dxixi} that it would be troublesome to refer to them each time. Therefore we kindly ask the reader to keep the formulae in mind since they will be used without any reference. 

We begin the calculations with the first term of the Hamiltonian \eqref{ham-pi-0}:
\begin{multline*}
G\Big(\pi^A-\frac{1}{2}(\vec{\theta}^C\lrcorner\pi_C){\theta}^A,\pi^B-\frac{1}{2}(\vec{\theta}^D\lrcorner\pi_D){\theta}^B\Big)=\\=
\Big([\pi^A-\frac{1}{2}(\vec{\theta}^C\lrcorner\pi_C){\theta}^A]\we{\theta}_B\Big)\we{*}\Big([\pi^B-\frac{1}{2}(\vec{\theta}^D\lrcorner\pi_D){\theta}^B]\we{\theta}_A\Big)-\\-\frac{1}{2}\Big([\pi^A-\frac{1}{2}(\vec{\theta}^C\lrcorner\pi_C){\theta}^A]\we{\theta}_A\Big)\we{*}\Big([\pi^B-\frac{1}{2}(\vec{\theta}^D\lrcorner\pi_D){\theta}^B]\we{\theta}_B\Big)=\\=(\pi^A\we{\theta}_B)\we{*}(\pi^B\we{\theta}_A)-\frac{1}{2}(\pi^A\we{\theta}_A)\we{*}(\pi^B\we{\theta}_B)-\\-(\vec{\theta}^C\lrcorner\pi_C){\theta}^A\we{\theta}_B\we{*}(\pi^B\we{\theta}_A)+\frac{1}{4}(\vec{\theta}^C\lrcorner\pi_C)^2{\theta}^A\we{\theta}_B\we{*}({\theta}^B\we{\theta}_A).
\end{multline*}
By virtue of \eqref{tt-bt} and \eqref{a-b} 
\[
-(\vec{\theta}^C\lrcorner\pi_C){\theta}^A\we{\theta}_B\we{*}(\pi^B\we{\theta}_A)=2(\vec{\theta}^C\lrcorner\pi_C)\pi_A\we*\theta^A=2(\pi_A\we{*}{\theta}^A)\we{*}(\pi_B\we{*}{\theta}^B)
\]
Applying \eqref{4t} in the first step, \eqref{a-b} and \eqref{a-b-e} in the second one we obtain
\[
\frac{1}{4}(\vec{\theta}^C\lrcorner\pi_C)^2{\theta}^A\we{\theta}_B\we{*}({\theta}^B\we{\theta}_A)=-\frac{6}{4}(\vec{\theta}^C\lrcorner\pi_C)^2\eps=-\frac{3}{2}(\pi_A\we{*}{\theta}^A)\we{*}(\pi_B\we{*}{\theta}^B).
\]
Thus
\begin{multline*}
G\Big(\pi^A-\frac{1}{2}(\vec{\theta}^C\lrcorner\pi_C){\theta}^A,\pi^B-\frac{1}{2}(\vec{\theta}^D\lrcorner\pi_D){\theta}^B\Big)=(\pi^A\we{\theta}_B)\we{*}(\pi^B\we{\theta}_A)-\\-\frac{1}{2}(\pi^A\we{\theta}_A)\we{*}(\pi^B\we{\theta}_B)+\frac{1}{2}(\pi_A\we{*}{\theta}^A)\we{*}(\pi_B\we{*}{\theta}^B).
\end{multline*}

Now let us introduce another map acting on pairs of one-forms $(\alpha^A,\beta^B)$: 
\begin{multline}
(\alpha^A,\beta^B)\mapsto \tilde{G}(\alpha^A,\beta^B):=G(\alpha^A,\beta^B)+\frac{1}{2}(\alpha_A\we{*}{\theta}^A)\we{*}(\beta_B\we{*}{\theta}^B)=G(\alpha^A,\beta^B)+\\+\frac{1}{2}\scal{\alpha^A}{\theta_A}\scal{\beta^B}{\theta_B}\eps\we*\eps=(q^{ik}q^{jl}-\frac{1}{2}q^{ij}q^{kl})\theta_{A(i}\alpha^A_{j)}\theta_{B(k}\beta^B_{l)}{\eps}
\label{tGAB}
\end{multline}
---here we used \eqref{g*g}, the fact that 
\begin{equation}
*\eps=1
\label{*e=1}
\end{equation}
and \eqref{GAB}. Note that $\tilde{G}$ is built from $(i)$ the same non-invertible linear mapping $\alpha^A_i\mapsto \theta_{A(i}\alpha^A_{j)}$ as $G$ and $(ii)$ the metric $\tilde{G}^{ij\,kl}:=(q^{ik}q^{jl}-\frac{1}{2}q^{ij}q^{kl})$ related to the metric $G^{ij\,kl}:=(q^{ik}q^{jl}-q^{ij}q^{kl})$ appearing in \eqref{GAB} as follows:
\[
\tilde{G}^{ij\,kl}G_{kl\,mn}=(q^{ik}q^{jl}-\frac{1}{2}q^{ij}q^{kl})(q_{km}q_{ln}-q_{kl}q_{mn})=\delta^i{}_m\delta^j{}_n.
\]     
Thus
\[
G\Big(\pi^A-\frac{1}{2}(\vec{\theta}^C\lrcorner\pi_C){\theta}^A,\pi^B-\frac{1}{2}(\vec{\theta}^D\lrcorner\pi_D){\theta}^B\Big)=\tilde{G}(\pi^A,\pi^B).
\]

Note now that $\pi^A$ in $\tilde{G}(\pi^A,\pi^B)$ undergoes the linear transformation $\pi^A_i\mapsto\theta_{A(i}\pi^A_{j)}$. According to \eqref{pi-df} $\pi^A$ contains the term $-\frac{1}{2}\vec{\theta}^A\lrcorner E^B{}_B$ which vanishes under the transformation:
\[
(\vec{\theta}^A\lrcorner E^B{}_B)_i=\theta^{Ak}E^B{}_{Bki}\mapsto\theta_{A(i}\theta^{Ak}E^B{}_{B|k|j)}=E^B{}_{B(ij)}=0.
\]      
Taking into account Equation \eqref{pi-rho} we see that
\[
G\Big(\pi^A-\frac{1}{2}(\vec{\theta}^C\lrcorner\pi_C){\theta}^A,\pi^B-\frac{1}{2}(\vec{\theta}^D\lrcorner\pi_D){\theta}^B\Big)=\tilde{G}(\rho^A,\rho^B)
\] 
and consequently \eqref{ham-pi-0} can be written as follows:
\begin{multline}
H_0[{\theta}^A,p_B,N,\vec{N}]=\int_\Sigma \frac{1}{2N}\Big(\tilde{G}(\rho^A,\rho^B)-E^A{}_B\we{*}E^B{}_A +\frac{1}{2}E^A{}_A \we{*}E^B{}_B\Big)+\\
+\frac{N}{2}F^A{}_B\we{*}F^B{}_A-\frac{N}{4}F^A{}_A\we{*}F^B{}_B.
\label{ham-rho}
\end{multline}

Our goal now is to express the terms
\begin{multline}
\tilde{G}(\rho^A,\rho^B)-E^A{}_B\we{*}E^B{}_A +\frac{1}{2}E^A{}_A \we{*}E^B{}_B=(\rho^A\we{\theta}_B)\we{*}(\rho^B\we{\theta}_A)-\\-\frac{1}{2}(\rho^A\we{\theta}_A)\we{*}(\rho^B\we{\theta}_B)+\frac{1}{2}(\rho_A\we{*}{\theta}^A)\we{*}(\rho_B\we{*}{\theta}^B)-E^A{}_B\we{*}E^B{}_A +\frac{1}{2}E^A{}_A \we{*}E^B{}_B
\label{rho-E}
\end{multline}
as explicite functions of the canonical variables, the lapse and the shift. To transform the five terms appearing at the r.h.s. of \eqref{rho-E} we express $\rho_A$, $E^A{}_B$ and $E^A{}_A$ as in \eqref{E-rho} and using repeatedly Equation \eqref{tt-eta}, the auxiliary formulae presented in Section \ref{aux} and Equation \eqref{*e=1} obtain in turn the first term:
\begin{multline*}
(\rho_A\we{\theta}^B)\we{*}(\rho_B\we{\theta}^A)=\\=\Big(N[{*}p_A-(\vec{\theta}_C\lrcorner{d}\xi^C){\theta}_A+{d}\xi_A]-(\vec{\theta}_C\lrcorner{\cal L}_{\vec{N}}{\theta}^C){\theta}_A+{\cal L}_{\vec{N}}{\theta}^C\vec{\theta}_C\lrcorner{\theta}_A\Big)\we{\theta}^B\we\\
\we{*}\Big[\Big(N[{*}p_B-(\vec{\theta}_D\lrcorner{d}\xi^D){\theta}_B+{d}\xi_B]-(\vec{\theta}_D\lrcorner{\cal L}_{\vec{N}}{\theta}^D){\theta}_B+{\cal L}_{\vec{N}}{\theta}^D\vec{\theta}_D\lrcorner{\theta}_B\Big)\we{\theta}^A\Big]=\\=
N^2\Big({*}p_A\we{\theta}^B\we{*}({*}p_B\we{\theta}^A)+2(\vec{\theta}^A\lrcorner{*}p_A)(\vec{\theta}^B\lrcorner{d}\xi_B){\eps}+2p_A\we{d}\xi^A-2(\vec{\theta}^B\lrcorner{d}\xi_B)^2{\eps}+\\+d\xi_A\we\theta^B\we*(d\xi_B\we\theta^A)\Big)+N\Big(4(\vth_A\lr\LL_{\vec{N}}\theta^A)(\vth^B\lr{*}p_B){\eps}+2{*}p_A\we\theta^B\we{*}(\LL_{\vec{N}}\theta_B\we\theta^A)-\\-6(\vth_A\lr{d}\xi^A)(\vth_B\lr\LL_{\vec{N}}\theta^B){\eps}+2{d}\xi_A\we{*}\LL_{\vec{N}}\theta^A\Big)-2(\vth_A\lr\LL_{\vec{N}}\theta^A)^2{\eps}+\LL_{\vec{N}}\theta_A\we\theta^B\we{*}(\LL_{\vec{N}}\theta_B\we\theta^A),
\end{multline*}
the second one:
\begin{multline*}
-\frac{1}{2}(\rho_A\we{\theta}^A)\we{*}(\rho_B\we{\theta}^B)=-\frac{1}{2}\Big(N[{*}p_A\we\theta^A+{d}\xi_A\we\theta^A]+\LL_{\vec{N}}\theta^A\we\theta_A\Big)\we\\\we{*}\Big(N[{*}p_B\we\theta^B+{d}\xi_B\we\theta^B]+\LL_{\vec{N}}\theta^B\we\theta_B\Big)=\\=N^2\Big(-\frac{1}{2}{*}p_A\we\theta^A\we{*}({*}p_B\we\theta^B)-{*}p_A\we\theta^A\we{*}({d}\xi_B\we\theta^B)-\frac{1}{2}({d}\xi_A\we\theta^A)\we{*}({d}\xi_B\we\theta^B)\Big)+\\+N\Big(-{*}p_A\we\theta^A\we{*}(\LL_{\vec{N}}\theta_B\we\theta^B)-({d}\xi_A\we\theta^A)\we{*}(\LL_{\vec{N}}\theta_B\we\theta^B)\Big)-\frac{1}{2}\LL_{\vec{N}}\theta_A\we\theta^A\we{*}(\LL_{\vec{N}}\theta_B\we\theta^B),
\end{multline*}
the third one:
\begin{multline*}
\frac{1}{2}(\rho_A\we{*}{\theta}^A)\we{*}(\rho_B\we{*}{\theta}^B)=\frac{1}{2}\Big(N[{*}p_A\we{*}\theta^A-2(\vth_C\lr{d}\xi^C){\eps}]-2(\vth_C\lr\LL_{\vec{N}}\theta^C){\eps}\Big)\we\\\we{*}\Big(N[{*}p_B\we{*}\theta^B-2(\vth_D\lr{d}\xi^D){\eps}]-2(\vth_D\lr\LL_{\vec{N}}\theta^D){\eps}\Big)=\\=
N^2\Big(\frac{1}{2}p_A\we\theta^A\we{*}(p_B\we\theta^B)-2(\vth^A\lr{*}p_A)(\vth_C\lr{d}\xi^C){\eps}+2(\vth_C\lr{d}\xi^C)^2{\eps}\Big)+\\+N\Big(-2(\vth^A\lr{*}p_A)(\vth_C\lr\LL_{\vec{N}}\theta^C){\eps}+4(\vth_C\lr{d}\xi^C)(\vth_D\lr\LL_{\vec{N}}\theta^D){\eps}\Big)+2(\vth_C\lr\LL_{\vec{N}}\theta^C)^2{\eps},
\end{multline*}
the fourth one:
\begin{multline}
-E^A{}_B\we{*}E^B{}_A=-\Big(-\xi^A{d}N\we{\theta}_B-N({d}\xi^A\we{\theta}_B-{d}{\theta}^A\we\xi_B)-{\cal L}_{\vec{N}}{\theta}^A\we{\theta}_B\Big)\we\\\we{*}\Big(-\xi^B{d}N\we{\theta}_A-N({d}\xi^B\we{\theta}_A-{d}{\theta}^B\we\xi_A)-{\cal L}_{\vec{N}}{\theta}^B\we{\theta}_A\Big)=-2N{d}N\we\theta_B\we{*}({d}\theta^B)+\\+N^2\Big(-{d}\xi^A\we\theta_B\we{*}({d}\xi^B\we\theta_A)-({d}\theta^A\we\xi_B)\we{*}({d}\theta^B\we\xi_A)\Big)+\\+N\Big(-2{d}\xi^A\we\theta_B\we{*}(\LL_{\vec{N}}\theta^B\we\theta_A)+2{d}\theta^A\we\xi_B\we{*}(\LL_{\vec{N}}\theta^B\we\theta_A)\Big)-\LL_{\vec{N}}\theta^A\we\theta_B\we{*}(\LL_{\vec{N}}\theta^B\we\theta_A)
\label{t-4}
\end{multline}
and, finally, the fifth one:
\begin{multline*}
\frac{1}{2}E^A{}_A \we{*}E^B{}_B=\frac{1}{2}\Big(2N\xi^Ad\theta_A-\LL_{\vec{N}}\theta^A\we\theta_A\Big)\we*\Big(2N\xi^Bd\theta_B-\LL_{\vec{N}}\theta^B\we\theta_B\Big)=\\=N^22\xi^Ad\theta_A\we*(\xi^Bd\theta_B)+N\big(-2\xi^Ad\theta_A\we*(\LL_{\vec{N}}\theta^B\we\theta_B)\big)+\frac{1}{2}\LL_{\vec{N}}\theta^A\we\theta_A\we*(\LL_{\vec{N}}\theta^B\we\theta_B).
\end{multline*}

Note now that each of the five terms consists of terms proportional to $N^2$, $N$ and ones which do not depend on $N$. Moreover, in \eqref{t-4} there is one term proportional to $NdN$. Let us now gather the corresponding terms obtaining thereby a decomposition of \eqref{rho-E} into terms proportional to $N^2,N, NdN$ and those independent of $N$. 

Gathering the terms we will try to simplify the formulae as much as possible. To this end we will also use the primary constraints \eqref{C-1} and \eqref{C-anti}---recall that at this moment we are still working with terms constituting the Hamiltonian $H_0$ which is defined only for $(\theta^A,p_B)$ satisfying the constraints.   

\paragraph{The term proportional to $N^2$} While gathering all the expressions proportional to $N^2$ we see that some terms cancel at once and we get
\begin{multline}
N^2\Big((*p_A\we\theta^B)\we*(*p_B\we\theta^A)-\frac{1}{2}(*p_A\we\theta^A)\we*(*p_B\we\theta^B)+\frac{1}{2}(p_A\we\theta^A)\we*(p_B\we\theta^B)+\\+2p_A\we{d}\xi^A-({*}p_A\we\theta^A)\we{*}({d}\xi_B\we\theta^B)-({d}\theta^A\we\xi_B)\we{*}({d}\theta^B\we\xi_A)+\\+\frac{3}{2}(\xi^Ad\theta_A)\we*(\xi^Bd\theta_B)\Big)
\label{N^2}
\end{multline}
To simplify this expression note first that the fifth term in \eqref{N^2} can be transformed as follows
\[
-({*}p_A\we\theta^A)\we{*}({d}\xi_B\we\theta^B)=-(\theta^A\we{*}p_A)\we{*}(\xi_Bd\theta^B)=-(\theta^A\we{*}p_A)\we*(\theta^B\we{*}p_B),
\]
where in the last step we used the constraint \eqref{C-anti}. This means that the fifth term is proportional to the second one. Because $\xi^A$ is a zero-form the sixth term in \eqref{N^2} is equal to 
\[
-(\xi_Ad\theta^A)\we{*}(\xi_B{d}\theta^B)
\] 
which means that it is proportional to the last term in \eqref{N^2}. By virtue of the constraint \eqref{C-anti} the sixth and the last terms are proportional to the second one. Thus the sum of the second, fifth, sixth and the last term reads
\[
-(*p_A\we\theta^A)\we*(*p_B\we\theta^B)
\] 
and consequently \eqref{N^2} can be written as
\begin{multline}
N^2\Big((*p_A\we\theta^B)\we*(*p_B\we\theta^A)-(*p_A\we\theta^A)\we*(*p_B\we\theta^B)+\frac{1}{2}(p_A\we\theta^A)\we*(p_B\we\theta^B)+\\+2p_A\we{d}\xi^A\Big).
\label{N^2-1}
\end{multline}
This expression can be simplified further---applying \eqref{atbt} to the first term of \eqref{N^2-1} and \eqref{atbt-1} to the second one we obtain 
\begin{multline*}
(*p_A\we\theta^B)\we*(*p_B\we\theta^A)-(*p_A\we\theta^A)\we*(*p_B\we\theta^B)=-(\vth^A\lr*p_A)(\vth^B\lr*p_B)\eps+\\+(\vth^A\lr*p_B)(\vth^B\lr*p_A)\eps=-(\theta^A\we p_A)\we*(\theta^B\we p_B)+(\theta^B\we p_A)\we*(\theta^A\we p_B),
\end{multline*}
where in the last step we used \eqref{a-b} and \eqref{a-b-e}. Thus we arrived at a simple form of the term in \eqref{rho-E} proportional to $N^2$:
\begin{equation}
N^2\Big((p_A\we\theta^B)\we*(p_B\we\theta^A)-\frac{1}{2}(p_A\we\theta^A)\we*(p_B\we\theta^B)+2p_A\we{d}\xi^A\Big).
\label{N^2-fin}
\end{equation}
 
\paragraph{The term proportional to $N$} Again while gathering all the expressions proportional to $N$ some terms cancel at once and we get 
\begin{multline}
N\Big(2(\vth_A\lr\LL_{\vec{N}}\theta^A)(\vth^B\lr{*}p_B){\eps}+2({*}p_A\we\theta^B)\we{*}(\LL_{\vec{N}}\theta_B\we\theta^A)-2(\vth_A\lr{d}\xi^A)(\vth_B\lr\LL_{\vec{N}}\theta^B){\eps}+\\+2{d}\xi_A\we{*}\LL_{\vec{N}}\theta^A -({*}p_A\we\theta^A)\we{*}(\LL_{\vec{N}}\theta_B\we\theta^B)-2({d}\xi^A\we\theta_B)\we{*}(\LL_{\vec{N}}\theta^B\we\theta_A)+\\+2({d}\theta^A\we\xi_B)\we{*}(\LL_{\vec{N}}\theta^B\we\theta_A)- (d\theta^A\we\xi_A)\we*({\cal L}_{\vec{N}}{\theta}^B\we{\theta}_B)\Big)
\label{N^1}
\end{multline}
Applying \eqref{atbt} to the second term of the expression above we see that the sum of the first and second terms can be expressed as
\[
2(*p_A)\we*(\LL_{\vec{N}}\theta^A)+2(\xi^A*p_A)\we*(\xi_B\LL_{\vec{N}}\theta^B)=2\LL_{\vec{N}}\theta^A\we p_A+2(\xi^A*p_A)\we*(\xi_B\LL_{\vec{N}}\theta^B)
\]  
On the other hand the seventh term in \eqref{N^1} can be transformed as follows
\begin{multline*}
2({d}\theta^A\we\xi_B)\we{*}(\LL_{\vec{N}}\theta^B\we\theta_A)=2(\xi_B\LL_{\vec{N}}\theta^B)\we\theta_A\we{*}{d}\theta^A=-2(\xi_B\LL_{\vec{N}}\theta^B)\we(\xi^Ap_A)=\\=-2(\xi_B\LL_{\vec{N}}\theta^B)\we*(\xi^A*p_A)=-2(\xi^A*p_A)\we*(\xi_B\LL_{\vec{N}}\theta^B)
\end{multline*}
---here in the second step we used the constraint \eqref{C-1}. Thus the sum of the first, the second and the seventh term is simply
\[
2\LL_{\vec{N}}\theta^A\we p_A.
\]
Moreover, the sum of the fifth and the last terms in \eqref{N^1} vanishes by virtue of the constraint \eqref{C-anti}: 
\begin{multline*}
-({*}p_A\we\theta^A)\we{*}(\LL_{\vec{N}}\theta_B\we\theta^B)- (d\theta^A\we\xi_A)\we*({\cal L}_{\vec{N}}{\theta}^B\we{\theta}_B)=-\\-(\theta^A\we*p_A-\xi_Ad\theta^A)\we*({\cal L}_{\vec{N}}{\theta}^B\we{\theta}_B)=0.
\end{multline*}
Thus we managed to simplify \eqref{N^1} to
\begin{multline*}
N\Big(2\LL_{\vec{N}}\theta^A\we p_A-2(\vth_A\lr{d}\xi^A)(\vth_B\lr\LL_{\vec{N}}\theta^B){\eps}+2{d}\xi_A\we{*}\LL_{\vec{N}}\theta^A -\\-2({d}\xi^A\we\theta_B)\we{*}(\LL_{\vec{N}}\theta^B\we\theta_A)\Big)
\end{multline*}
Now it is enough to apply \eqref{atbt} to the last term of the expression above to realize that \eqref{N^1} reduces to
\begin{equation}
2N\LL_{\vec{N}}\theta^A\we p_A
\label{N^1-fin}
\end{equation}
which is the final expression of the terms in \eqref{rho-E} proportional to $N$. 

\paragraph{The term independent of $N$}
Gathering appropriate terms we obtain
\begin{multline*}
-2(\vth_A\lr\LL_{\vec{N}}\theta^A)^2{\eps}+\LL_{\vec{N}}\theta_A\we\theta^B\we{*}(\LL_{\vec{N}}\theta_B\we\theta^A)-\frac{1}{2}\LL_{\vec{N}}\theta_A\we\theta^A\we{*}(\LL_{\vec{N}}\theta_B\we\theta^B)+\\+2(\vth_C\lr\LL_{\vec{N}}\theta^C)^2{\eps}-\LL_{\vec{N}}\theta^A\we\theta_B\we{*}(\LL_{\vec{N}}\theta^B\we\theta_A)+\frac{1}{2}{\cal L}_{\vec{N}}{\theta}^A\we{\theta}_A\we*({\cal L}_{\vec{N}}{\theta}^B\we{\theta}_B)=0
\end{multline*}

In this way we managed to simplify \eqref{rho-E} to a sum of the term in \eqref{t-4} proportional to  $NdN$ and the expressions \eqref{N^2-fin} and \eqref{N^1-fin}:
\begin{multline}
\tilde{G}_{AB}(\rho^A,\rho^B)-E^A{}_B\we{*}E^B{}_A +\frac{1}{2}E^A{}_A \we{*}E^B{}_B=-2NdN\we\theta_B\we*d\theta^B+\\+N^2\Big((p_A\we\theta^B)\we*(p_B\we\theta^A)-\frac{1}{2}(p_A\we\theta^A)\we*(p_B\we\theta^B)+2p_A\we{d}\xi^A\Big)+\\+2N\LL_{\vec{N}}\theta^A\we p_A.
\end{multline}
Setting this to the Hamiltonian \eqref{ham-rho} we obtain
\begin{multline}
H[{\theta}^A,p_B,N,\vec{N}]=\int_\Sigma -dN\we\theta_B\we*d\theta^B +N\Big(\frac{1}{2}(p_A\we\theta^B)\we*(p_B\we\theta^A)-\\-\frac{1}{4}(p_A\we\theta^A)\we*(p_B\we\theta^B)+p_A\we{d}\xi^A+\frac{1}{2}F^A{}_B\we{*}F^B{}_A-\frac{1}{4}F^A{}_A\we{*}F^B{}_B\Big)+\\+\LL_{\vec{N}}\theta^A\we p_A.
\label{ham-expl-0}
\end{multline}

What remains to be done is to remove the derivatives of the laps $N$ and the shift $\vec{N}$ appearing, respectively, in the first and in the last terms of the Hamiltonian above.
Applying the constraint \eqref{C-1} to the first term we obtain
\begin{multline*}
-dN\we\theta_B\we*d\theta^B=dN\we\xi^Ap_A=d(N\xi^Ap_A)-N(d\xi^Ap_A)=\\=
d(N\xi^Ap_A)-Nd\xi^A\we p_A-N\xi^A\we dp_A.
\end{multline*}
On the other hand it was shown in \cite{os} that 
\[
({\cal L}_{\vec{N}}{\theta}^A)\we p_A=-{d}{\theta}^A\we(\vec{N}\lrcorner p_A)-(\vec{N}\lrcorner{\theta}^A)\we {d}p_A+{d}((\vec{N}\lrcorner{\theta}^A)\we p_A)
\]
Since $\Sigma$ is a compact manifold without boundary the terms $d(N\xi^Ap_A)$ and ${d}((\vec{N}\lrcorner{\theta}^A)\we p_A)$ vanish once integrated over $\Sigma$. Rewriting $F^A{}_B$ as a function of $\theta^A$ (see \eqref{F}) we arrive finally at the Hamiltonian $H_0$ expressed explicitely as a function of the canonical variables, the laps and the shift
\begin{multline}
H_0[{\theta}^A,p_B,N,\vec{N}]=\int_\Sigma N\Big(\frac{1}{2}(p_A\we\theta^B)\we*(p_B\we\theta^A)-\frac{1}{4}(p_A\we\theta^A)\we*(p_B\we\theta^B)-\\-\xi^A\we{d}p_A+\frac{1}{2}(d\theta_A\we\theta^B)\we{*}(d\theta_B\we\theta^A)-\frac{1}{4}(d\theta_A\we\theta^A)\we{*}(d\theta_B\we\theta^B)\Big)-\\-{d}{\theta}^A\we(\vec{N}\lrcorner p_A)-(\vec{N}\lrcorner{\theta}^A)\we {d}p_A,
\label{ham-expl}
\end{multline}
which is exactly the Hamiltonian \eqref{H_0}. In order to extend $H_0$ to the whole phase space we add to it the smeared primary constraints \eqref{B-sm} and \eqref{R-sm} and arrive thereby at \eqref{full-ham}.  

The Hamiltonian \eqref{full-ham} depends on the Lagrange multipliers $N$ and $\vec{N}$. Variations of the Hamiltonian with respect to the multipliers give us the secondary constraints \eqref{sec-constr}. Expressing the r.h.s. of \eqref{H_0} and \eqref{full-ham} by means of the smeared versions \eqref{S-sm} and \eqref{V-sm} of, respectively, the scalar and the vector constraints gives us \eqref{H0-SV} and \eqref{full-ham-c}.

\paragraph{Acknowledgments} I am grateful to J\k{e}drzej \'Swie\.zewski for his cooperation in the research on the Hamiltonian framework of YMTM described in \cite{os} which was for me a preparatory exercise for deriving the results described in this paper and in \cite{oko-tegr-II}. I am also grateful to {prof. Jerzy Kijowski and prof. Pawe{\l} Nurowski for discussions}, to prof. Jerzy Lewandowski and prof. Jacek Jezierski for useful comments which allowed me to simplify at some points the results and the presentation of them and {to a reviewer for pointing out to me the work \cite{wall-av} by Wallner I was not aware of while deriving the results presented in this paper.}  



\end{document}